**Title:** Search for structural differences in spike glycoprotein variants of SARS-CoV-2: Infrared Spectroscopy, Circular Dichroism and Computational Analysis.

*Tiziana Mancini[a,*], Nicole Luchetti[b], Salvatore Macis[a], Velia Minicozzi[c], Rosanna Mosetti[d], Alessandro Nucara[a], Stefano Lupi[a] and Annalisa D'Arco[a,*]*

[a] Department of Physics, University La Sapienza, P.le A. Moro 2, 00185, Rome, Italy
[b] Engineering Department, Università Campus Bio-Medico di Roma, Via Alvaro del Portillo 21, 00128, Rome, Italy
[c] Department of Physics, University of Rome Tor Vergata, Via della Ricerca Scientifica, 1 00133 Rome, Italy
[d] Department of Basic and Applied Sciences for Engineering (SBAI). Sapienza University of Rome, Via A. Scarpa 16, 00161, Rome, Italy

**Correspondence**
Tiziana Mancini, Annalisa D'Arco
Department of Physics, University La Sapienza, P.le A. Moro 2, 00185, Rome, Italy
E-mail: tiziana.mancini@uniroma1.it, annalisa.darco@uniroma1.it

**Abstract:** The SARS-CoV-2 pandemic has led to a significant emergence of highly mutated forms of viruses with a great ability to adapt to the human host. Some mutations resulted in changes in the amino acid sequences of viral proteins, including the Spike glycoproteins, affecting protein physico-chemical properties and functionalities. Here, we propose, for the first time to the best of our knowledge, a systematic and comparative study of the monomeric spike protein subunits 1 of three SARS-CoV-2 variants at pH 7.4, combining both an experimental approach, taking advantage of Attenuated Total Reflection Infrared and Circular Dichroism spectroscopies, and a computational approach via Molecular Dynamics simulations. Experimental data in combination with Molecular Dynamics and Surface polarity calculations provide a comprehensive understanding of variants' proteins in terms of their secondary structure content, 3D conformational structure and order and interaction with the solvent. The present structural investigation clarifies which kind of changes in conformation and functionalities occurred as long as mutations appeared in amino acids sequences. This information is essential for preventive targeted actions, drug design, and biosensing applications.

## 1. Introduction

SARS-CoV-2 pandemic, which arose in 2019, has been the largest-scale health emergency of the last few centuries, causing more than 7 million deaths to date, as declared by the World Health Organization (WHO) (1,2). SARS-CoV-2 virus responsible for this pandemic crisis is a member of the species severe acute respiratory syndrome-related Coronaviridae family. Similarly to other Coronaviruses, SARS-CoV-2 shows a spherical enveloped structure (with a diameter between 80-120 nm) enclosing a single positive strand RNA. Besides the different proteins included in the virus structure (Envelope, Membrane, Nucleocapsid and Spike proteins), the Spike (S) glycoprotein plays the most important role in the viral transmission process. S glycoprotein is the largest viral protein in the Coronavirus family, made up of 1273 amino acids and protruding from the viral surface with the task to anchor to the human receptor ACE2 (3). S protein is composed of two subunits, S1 (from 1-685 amino acids (aa)) and S2 (686-1273 aa), each one formed by different domains. In turn, S1 subunit is composed of the N-terminal domain (NTD, 14–305 aa) and the receptor binding domain (RBD, 319–541 aa): the latter is the protein site that first binds to ACE2 and initiates the infectious process. (4-9).

The virus circulation for almost a year resulted in a major step forward in its adaptation to humans (10,11). Numerous mutations occurred in the viral RNA genomic sequences over time, giving rise to highly mutated forms of SARS-CoV-2 which have been recognized and classified into viral lineages named Variants Of Concern (VoCs) (12-15). Some mutations led to variations in the amino acid sequences of viral proteins, such as single amino acid substitution, amino acids deletion or insertion, also resulting in structural changes of S glycoproteins. Moreover, single-point mutations can even lead to marked functional changes, despite causing only minor alterations in the overall structure. Such mutations may affect local flexibility, disrupt or create hydrogen bonds, modify the electrostatic potential, or alter side-chain packing, which in turn can impact protein–protein interactions, stability, or receptor binding. These seemingly small perturbations can have long-range allosteric effects or shift the equilibrium between functional states. Many works on SARS-CoV-2 VoCs have documented the influence of mutations on the S protein behavior, especially in terms of interaction bridges and strength with ACE2 receptor, binding affinity (16-19), flexibility (20,21), accessibility (14,22). These findings point out possible differences and changes in protein properties, and as a consequence, how mutations affect the evolution and spread of the virus (6,16, 20-31). For instance, it is well known that the Omicron variant exhibits significantly improved transmissibility (32-35), with 50 mutations compared to the wild type (WT). Twenty-six of these mutations are unique to this variant and more than 30 concern the spike glycoprotein. Therefore, the success of each VoC relative to the previously dominant one was enabled by altered intrinsic functional properties of the virus and, to varying degrees, by changes to virus antigenicity, conferring the ability to evade a primed immune response (12,17, 36,37). In this context, the knowledge of the secondary structural characteristics of SARS-CoV-2 VoCs protein and their differences are of primary importance to understand the connection with the spreading and infectious mechanisms.

Our work provides for the first time, to the best of our knowledge, a systematic and comparative study of three S1 monomeric subunits of VoCs of SARS-CoV-2 virus. Both experimental and computational approaches have been employed, combining infrared (IR) and circular dichroism (CD) spectroscopies, which are well-established experimental methods for a non-invasive analysis of polypeptides and proteins (37-45), with Molecular Dynamics (MD) simulations. In this context, our study represents a fundamental step in the non-invasive and non-destructive investigation and comparison of three dominant VoCs, that affected Europe and the world during the period 2020-2022. Referring to them by their common names, we considered Alpha variant (lineage B.1.1.7), Gamma variant (lineage P.1/P.1.1/P.1.2) and Omicron variant (lineage B.1.1.529).

The analysis of possible effects of VoCs mutations on protein structure is firstly provided by Circular Dichroism (CD) spectra in the wavelength range between 190-230 nm, whose signal arises from the π-π * and n-π * electronic transitions of the peptide group. These transitions are strongly related to the chiral properties of secondary structures, in terms of α-helix, β-sheet, β-turn and random coil structures (40), making CD spectroscopy an effective tool for providing a preliminary visualization of structural changes resulting from viral mutations. A subsequent in-depth analysis of secondary structure has been performed exploiting IR spectroscopy and focusing the study on amide I band (1590-1710 cm$^{-1}$), through an analysis of its vibrational spectral components (37-39). IR and CD spectra are also strongly influenced by the general three-dimensional (3D) conformational structure of the protein and therefore, its intrinsic conformational order, in terms of bonds and angles symmetry (37, 46-48), by its packaging and folding, its hydrophilicity and finally by the strength of hydrogen bonds with surrounding water (49,50). Spectral differences of Alpha, Gamma and Omicron S1 proteins, detected through IR and CD spectra, are accompanied by the description of their behavior in water obtained through a computational approach employing ColabFOLD, MD simulations and Protein-sol software. Through this combined analysis, a deep insight into the conformational structure of the three VoCs was achieved, pointing out important differences in their structural dynamics, from both experimental measurements and computational simulations. Our overall results, achieved for Alpha, Gamma and Omicron S1 proteins, help to shed light on the different VoCs functionalities and properties. In particular, the knowledge of their secondary structure and conformational behavior in water sheds light on their functionalities and the structural dynamics underlying the infectious mechanism. Therefore, these results constitute a pillar for several applicative fields, from the design of innovative optical biosensors, where viral proteins may be used as a potential biomarker (51-57), to drug design, preventive action development and further structural studies.

## 2. Materials and Methods

### 2.1 Protein preparation

Monomeric S1 subunits of S glycoprotein from three different variants of SARS-CoV-2 virus have been considered. Alpha variant S1, lineage B.1.1.7 (Cat. No. 40591-V08H12, aa 678, purity > 90 %), Gamma variant S1, lineage P.1/P.1.1/P.1.2 (Cat. No. 40591-V08H14, aa 681, purity > 90 %) and Omicron variant S1, lineage B.1.1.529 (Cat. No. 40591-V08H41, aa 678, purity > 95 %) have been all purchased from Sino Biological Europe GmbH (Eschborn, Germany). They were expressed in baculovirus insect cells with the same purity > 90% as determined by sodium dodecyl sulphate–polyacrylamide gel electrophoresis (SDS-PAGE) and finally used without further purification. They differ from the S1 protein of SARS-CoV-2 wild type (WT) (Cat. No. 40591-V08B1, aa 681, purity > 90%), referred to the variant that affected Wuhan in 2019, for a number 7,10 and 31 of mutations, respectively. Their amino acid sequences are reported in SI (paragraph S1), and the mutations characterizing them are schematized in Table 5. Common mutations are reported with the same colors.

**Table 5.** Variants of Concern (VoCs) and their mutations. Their standard nomenclature, lineages and mutations with respect to Wuhan wild type (WT) virus S1 protein (Cat. No. 40591-V08B1) are listed. Legend: A(xyz)B means that amino acid A in position (xyz) has been substituted with amino acid B. A(xyz)del means that amino acid A in position (xyz) has been deleted. ins(xyz)ABC means that the insertion of ABC amino acids in position (xyz) has occurred.

| Variant | Lineage | Mutations |
|---------|---------|-----------|
| Alpha | B.1.1.7 | H69del  V70del  Y144del  N501Y<br>A570D  D614G  P681H |
| Gamma | P.1/P.1.1/P.1.2 | L18F  T20N  P26S  D138Y  R190S |
| | | K417T  E484K  N501Y  D614G  H655Y |
| Omicron | B.1.1.529 | A67V  H69del  V70del  T95I  G142D  V143del  <br> Y144del  Y145del  N211del  L212I  ins214EPE  <br> G339D  S371L  S373P  S375F  K417N  N440K  <br> G446S  S477N  T478K  E484A  Q493R  G496S  <br> Q498R  N501Y  Y505H  T547K  D614G  H655Y  <br> N679K  P681H |

Each lyophilized protein was reconstructed by dissolving 100 µg of the pellet in distilled water (400 µL, pH 7.4) obtaining solutions with a concentration of 0.25 mg/ml. Other concentrations were investigated in our previous work (58,59), obtaining similar data and verifying the independence of IR measurements from concentrations.

## 2.2 Attenuated-Total-Reflection Infrared Spectroscopy and Data Analysis

ATR-IR spectra of monomeric S1 protein of the three SARS-CoV-2 VoCs were collected using a Bruker (Billerica, MA, USA) Vertex 70v Michelson spectrometer integrated with an ATR–Diamond single reflection module and a DLaTGS wide range detector. IR Spectroscopic measurements were performed at room temperature (25°C) and under vacuum conditions, in order to mitigate the interferences induced by water vapour and CO2 absorptions. The background spectrum (water solvent) was collected immediately prior to each sample measurement, with the same experimental settings of the protein solution. For each ATR-IR measurement 5 µl of the solution (protein solution or water one) were directly dropped on the ATR diamond crystal, collecting seven repetitions of 128 scans between 400–4000 $cm^{-1}$ with a spectral resolution of 2 $cm^{-1}$. For each protein sample, five independent depositions were measured and analysed. The ATR crystal was cleaned with ethanol (purity > 90 %), distilled water and subsequently with a lens tissue in order to eliminate any spurious signal. Spectral analysis (absorbance calculation, baseline correction, water subtraction, ATR advanced correction, cut and average) have been performed using both OPUS 8.2. software (Bruker Optics) and algorithms based on MATLAB (ver. 2018, MathWorks Inc., Natick, MA, USA), as described in our previous works (58,59). Protein secondary structures were studied focusing on the Amide I vibrational absorption band (37-39), lying in the spectral range from 1590-1720 $cm^{-1}$. For each protein, ATR-IR spectrum of the Amide I band was normalized to the maximum value and deconvoluted into its spectral components. The frequencies achieved by 2nd-derivative absorption spectra were used as starting points for multiple gaussian fitting performed with OPUS 8.2 software considering residual error (RMSE) value as goodness of fit parameter (55,56,58,59). The area of each convoluted band was normalized to the total integrated intensity after the subtraction of side chain vibrations contribution (60) and was used to calculate the percentage of each absorption band and then to estimate proteins secondary structures percentage content (60-63,86). The error associated with each secondary structure percentage content is calculated propagating the standard deviations of the convoluted band integrals percentage contribution obtained for each protein measurement run by adapting the final fit on its spectrum (61).

*2.3 Circular Dichroism Spectroscopy and Data Analysis*

To analyze the chirality of monomeric S1 proteins through their optical activity, CD spectra were collected with J715 Spectropolarimeter by Jasco (Tokyo, Japan), equipped with a high intensity Xenon lamp, emitting in the UV region from 180 nm to 700 nm. An oscillating linear crystal creates circularly polarized light, and a photomultiplier tube converts the light signal in current. A constant flow of nitrogen gas is used inside the spectropolarimeter in order to prevent the interferences of vapour water components. Samples are placed in a quartz suprasil cuvette (Hellma, Germany) with the optical path of 0.01 nm and it is thermalized at 25 °C with a Heto S.r.l. (Italy) Refrigerated Circulation Bath CBN 8-30. Ethanol (CAS 64-17-5, Carlo Erba, purity > 90 %) and distilled water are used to clean the cuvette. For each sample, the CD spectrum is the average of three independent CD measurements (meaning the cuvette has been filled with protein solution and washed three times), collected in step scanning mode, from 190 nm to 240 nm, with a 0.5 nm data pitch, bandwidth of 1 nm, and the exposure time of 8 s. Raw data (measured in mdeg) are converted to differential absorption coefficient ($\Delta\varepsilon$) normalized on the amino acids number (according to the equation reported in S2). The instrument interfaces through the Jasco Spectra Manager™ software which has been used for the first spectra processing (background subtraction, average and unit conversion). For the deconvolution, CD spectra fits are performed with CDpro software package (https://sites.google.com/view/sreerama) (64). This approach has been selected since CDPro is commonly employed for membrane proteins analysis and for the estimation of secondary structure content (65,66). Datasets were selected according to the type of proteins of interest (globular and membrane), and to the frequency range where spectra have been collected. It contains four algorithms for CD spectra analysis, namely CONTINLL, CDSSTR, SELCON3 and CLUSTR algorithms, and it can refer to 10 different datasets. For the deconvolution of S1 proteins CD spectra, both CONTINLL, CDSSTR and SELCON3 algorithms were employed, with different reference basis sets, generally used for membrane and soluble proteins, namely SP37, SP43, SDP42, SDP48, SMP50 and SMP56.

For each S1 protein CD spectrum, final fit is calculated as the average of fitting spectra obtained from each of the algorithms employing each of the basis sets. Also, secondary structure percentage contents were estimated as the average and standard deviation of percentage content obtained from each of the algorithms employing each of the basis sets (65, 66).

*2.4 ColabFold, Molecular dynamics simulation and Protein-sol software*

ColabFold software is employed to compute three-dimensional (3D) structure prediction of S1 proteins starting from their known FASTA amino acid sequences, outperforming AlphaFold2 (67) algorithm with Many-against-Many sequence searching (MMseqs2) server (68,69). Structure files in pdb format were produced and visualized with PyMOL (https://pymol.org/).

Molecular Dynamics (MD) simulations and post-processing analyses were performed with the GROMACS v. 2022.3 package (63, 68), using models obtained from ColabFold predictions as starting points for each variant. The centre of mass of each protein is placed in the centre of a cubic box of dimensions such that nearby images lay 10 Å away. The box is filled with TIP3P water molecules and 0.15 M of NaCl to make the whole system neutral. CHARMM and AMBER are among the most widely used force fields for protein simulations. The CHARMM force field, in particular, has been employed in several studies involving the Spike protein (70,71). While CHARMM36 could be used, especially given its suitability for intrinsically disordered proteins, we opted for CHARMM22/CMAP to model all proteins due to its favourable performance in previous studies focused on vibrational and IR spectral properties of proteins. Moreover, this force field is able to reproduce the transitions between α-helix and β-sheet secondary structures and to maintain a precise balance between α-helix and random coil conformations, facilitated by the inclusion of a supplementary term in the dihedral potential, known as the CMAP term (69, 72-74). Additionally, the CHARMM22* force field incorporates the Urey-Bradley term (69, 74, 75), which accounts for angle bending through 1,3 nonbonded interactions. This feature enhances the accuracy of molecular vibration assessments compared to other widely employed force fields.

For all simulation systems, firstly a minimization phase was performed, for $5 \cdot 10^3 + 5 \cdot 10^3$ steps of steepest descent and conjugate gradient algorithms in series, with a maximum force value of 10 kJ·mol$^{-1}$·nm$^{-1}$. To reduce the degrees of freedom, constraints on all the hydrogen bonds are imposed with the LINCS algorithm (76) for all the simulation phases. The equilibration strategy adopted consists of i) 3 × 20 ns in the NVT ensemble at different temperatures (150, 200 and 300 K), with position restraints on the protein backbone and side chains to relax the solvent around the protein, and ii) 20 ns in the NpT ensemble without restraints. After the equilibration, 600 ns production simulations were performed in the NpT ensemble, for a single replica of each model. The temperature of the whole system was kept fixed at room temperature using the V-rescale thermostat (77), with a coupling time of 0.1 ps, and the pressure was kept fixed at 1 bar using the Parrinello-Rahman barostat (78,79) with a coupling time of 2 ps and an isothermal compressibility of $4.5 \cdot 10^{-5}$ bar$^{-1}$. We employed the Particle Mesh Ewald algorithm (80) to handle the Coulomb interaction. A time step of 2 fs and a non-bonded pair list cut-off of 1.0 nm were used. The list was updated every 10 steps. The analysis of the numerical data obtained from the simulations was carried out using GROMACS and Python v. 3 (81) handmade programs, together with Visual Molecular Dynamics v. 1.9.3 tool (82) for trajectory visualization and analysis, according to the needs. FES heatmaps are computed in two dimension and calculated for the RMSD and the Rg. The color scale corresponds to the free energy regarding the probability of the possible pairs of values for the quantities under consideration. The free energy is defined in correlation with the canonical partition function (in the logarithmic term) in the following way:

$$G = -0.001 \cdot Av \cdot Kb \cdot T \cdot (\log_{10}(Z) - \log_{10}(\max(Z)))$$

G is expressed in kcal/mol, with Kb = 3.2976268E-24 (cal/K) that is the Boltzmann constant, Av = 6.0221417923 is the Avogadro number, and T = 298 (K) the temperature. The square matrix Z contains the frequency of occurrence of the pairs of RMSD and Rg values.

Protein-sol software (https://protein-sol.manchester.ac.uk/) was employed to compute hydrophobicity patches on S1 proteins surface (83). In particular, for each variant protein, non-polar to polar (NPP) ratio surface (84) was computed.

### 3. Results

Monomeric S1 proteins from Alpha, Gamma and Omicron SARS-CoV-2 variants have been modelled with AlphaFold2 algorithm, outperformed by ColabFold (57), starting from their amino acid sequences, provided by Sino Biological Europe GmbH, and reported in Supplementary Information (SI) (paragraph S1). In Figure 1, three-dimensional models of the three S1 proteins are visualized with PyMOL (https://pymol.org/) and name and position of each mutation are pointed out. RBDs are highlighted in darker color for each variant protein.

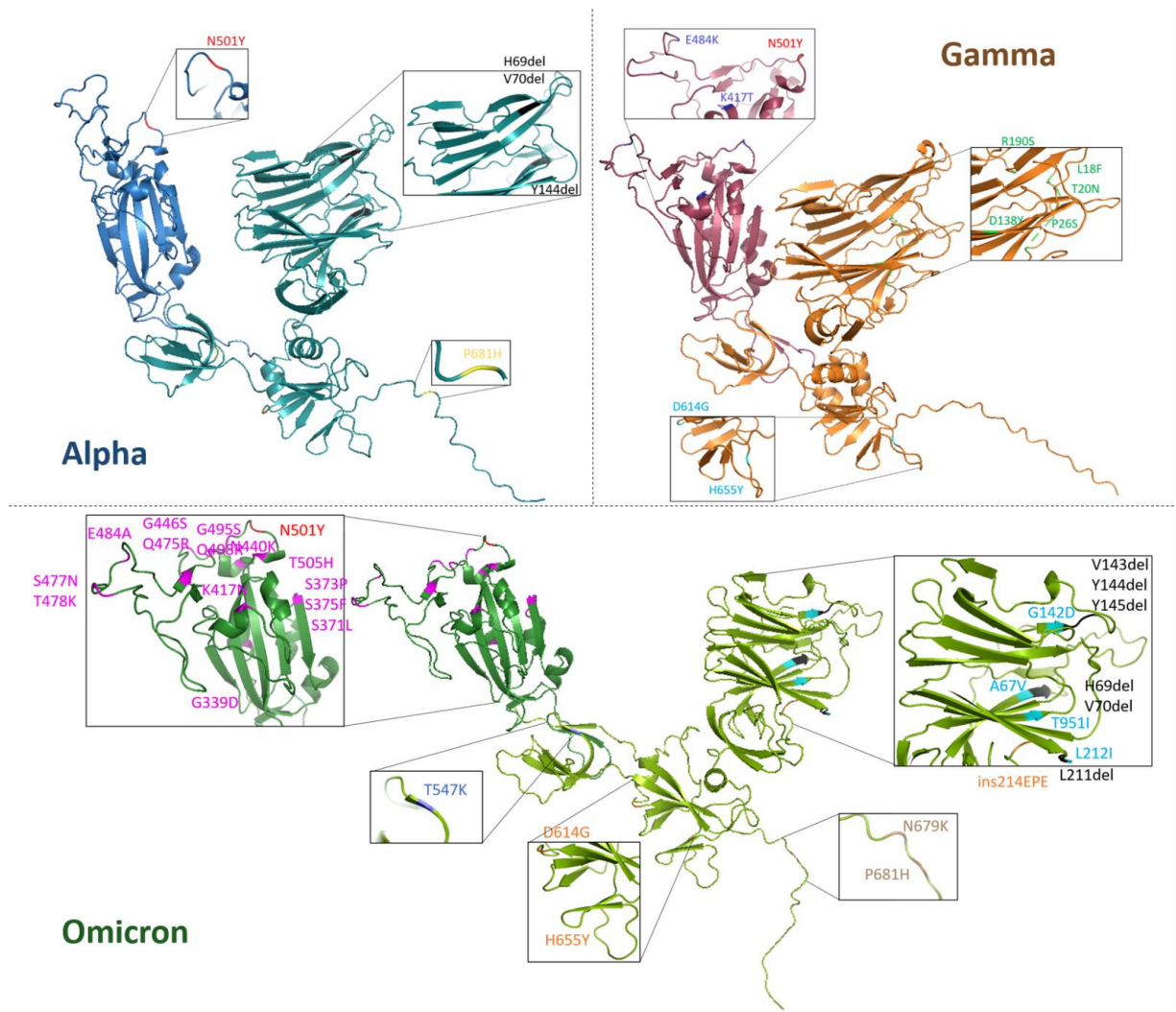

**Figure 1**. 3D Visualization of monomeric S1 protein variants. Alpha (blue), Gamma (orange) and Omicron (green) SARS-CoV-2 variants. For each S1 protein, RBD region (319-541 aa) is highlighted with a darker color with respect to the whole protein. For each variant S1 protein, respective mutations are zoomed and listed in the inset. The common mutation N501Y is indicated with the same color (red) and the deletions are marked in black.

An in-depth spectral analysis was performed for the three different SARS-CoV-2 variants, using both CD and IR spectroscopy, in order to investigate possible different structures among S1 proteins which in turn have an extremely similar primary structure (see Pairwise Sequence Alignment Emboss Needle pdf files in SI). Spectral differences were recognized and interpreted, bringing them back to differences in the 3D S1 conformational structures, and therefore on their functionalities and interaction with the solvent.

At the same time, AlphaFold2 modeling, protein visualization in 3D space, MD simulations and protein-sol software were employed to provide a comprehensive picture of the structural and physico-chemical features of the three proteins and to verify the changes induced by mutations as the virus evolves and adapts to the environment.

### 3.1 CD spectral analysis

CD spectra (expressed in terms of ellipticity variation Δε (40) were collected for the S1 proteins of SARS-CoV-2 variants (Alpha, Gamma and Omicron) between 190 and 230 nm. In this spectral region, the protein absorption is mainly due to peptide bond with its n-π* transition (around 210 nm) and π-π* transition (around 190 nm) (40-42). CD spectra (colored lines) are shown in Figure 2A for Alpha, Gamma and Omicron S1 proteins. Noticeable differences occur in the CD spectra of the three S1 protein variants, proving that mutated proteins are structurally different from each other.

A spectral variation is noticeable at the lowest wavelengths (190 nm), where CD absorptions of Gamma and Omicron S1 proteins have a more intense and broader peak (centered at about 194 nm) compared to the Alpha one. In this wavelength range, both the positive dichroic absorption of $\beta$-sheets structures and the negative absorption of $\beta$-turn structures give their competitive contribution (40).

An evident variation is also visible between 200 and 220 nm, where the negative absorption contribution is due to both β-sheets and α-helix structures. In particular, the chirality of α-helices gives rise to the characteristic modulation, which appears as two negative and partially overlapping peaks (40). These peaks are clearly distinguishable for each spectrum, being located at 211 nm and 215 nm for Alpha S1 protein, at 210.5 nm and 215 nm for Gamma S1 and at 212 nm and 215 nm for Omicron S1. The variations of $\beta$-sheets and $\alpha$-helices contribution in Alpha, Gamma and Omicron proteins appear also at the edge of the negative dip between 218 and 240 nm, where the three CD spectra clearly show different behaviors. In particular, in this frequency interval, the CD spectrum of S1 Omicron protein (green curve) presents lower values of Δε (calculated as reported in S2) compared to Alpha and Gamma CD spectra (blue and orange curve in Figure 2A, respectively). It is worth mentioning that the WT S1 protein has been measured, too, in order to observe the extent of structural change starting from the first virus and going on as long as mutations occur. All four of the CD spectra are comparatively reported in SI (see S3, Figure S1). Significant spectral variations are visible at the low wavelengths (190 nm) and at higher wavelengths (between 200-230 nm), testifying to an important effect of mutations on variants protein structure.

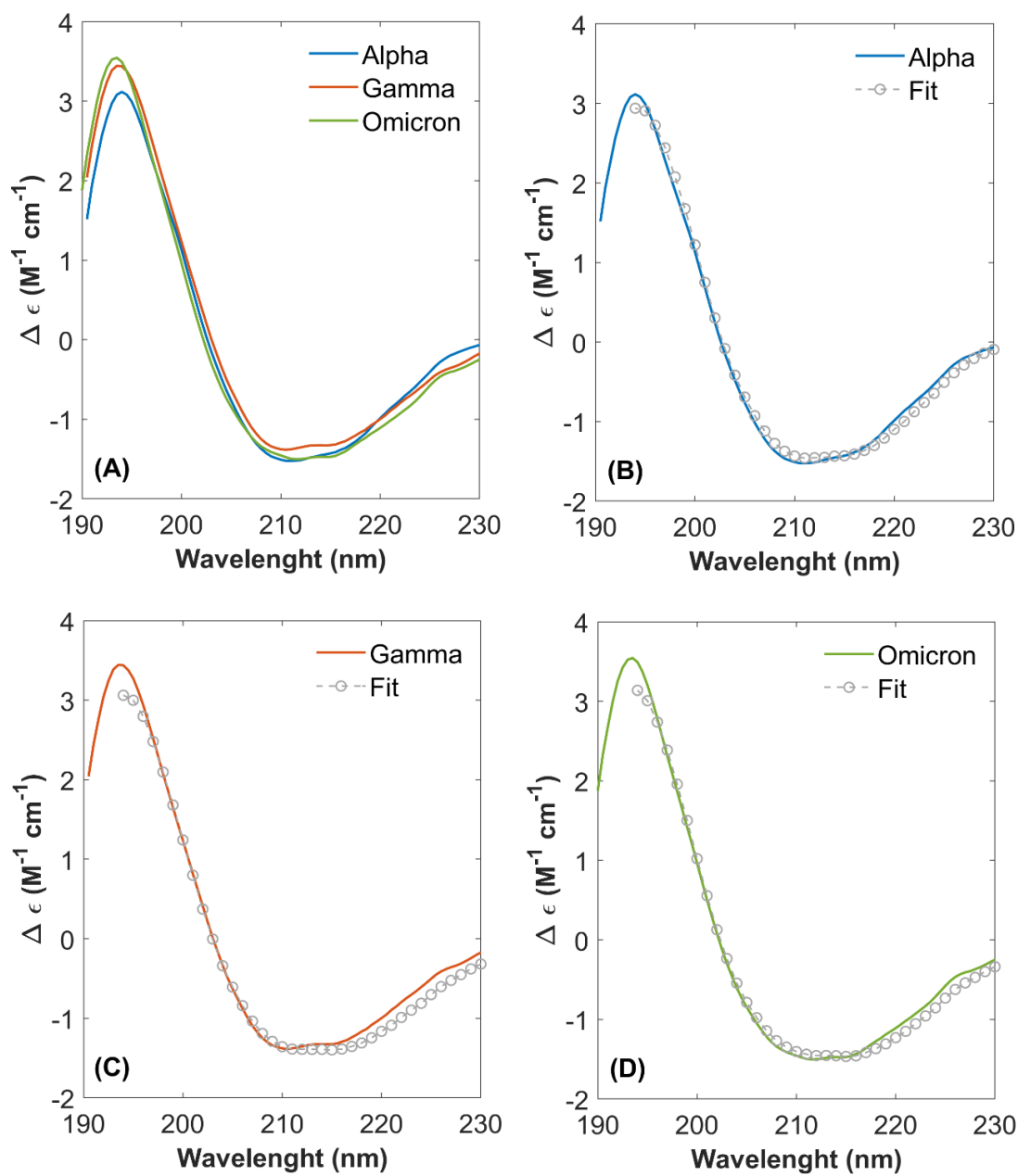

**Figure 2.** Comparison of S1 CD spectra between 190 and 230 nm. (A) Overlap and direct comparison of CD spectra between 190 and 230 nm of S1 proteins from SARS-CoV-2 Alpha (blue), Gamma (orange) and Omicron (green) variants. Single variant spectra (colored curves) with fitting curves (grey circles), for (B) Alpha, (C) Gamma and (D) Omicron proteins.

CDPro software was used to perform the CD spectra deconvolution of the three proteins from 195 to 230 nm. We selected this wavelength range because it is the largest common interval where all the spectral databases employed for the CD analysis efficiently work. Final fitting curves are displayed in Figure 2B, 2C and 2D (grey circles) for Alpha, Gamma and Omicron S1 protein spectra, respectively, as explained in the Materials and Methods section. From the deconvoluted analysis, the percentage content of the secondary structure was deduced and is reported in Table 1 for each variant

S1 proteins, as the MEAN ± SD calculated over the results obtained from the three algorithms and the six reference basis sets. See details for analysis in the Materials and Methods section.

Table 1. Secondary structure percentage content from CD analysis. Percentage fractions of the secondary structure of monomeric S1 protein of SARS-CoV-2 Alpha, Gamma and Omicron variants are estimated from CD spectra deconvolution performed with CDSSTR, SELCON3 and CLUSTR algorithm using SP37, SP43, SDP42, SDP48, SMP50 and SMP56 basis sets.

|  | $\beta$-sheet (%) | $\alpha$-helix (%) | $\beta$-turn (%) | Random coil (%) |
|---|---|---|---|---|
| Alpha | 38 ± 7 | 8 ± 4 | 23 ± 7 | 31 ± 7 |
| Gamma | 39 ± 6 | 11 ± 6 | 20 ± 2 | 30 ± 4 |
| Omicron | 37 ± 7 | 8 ± 4 | 23 ± 3 | 32  5 |

### 3.2 IR spectral analysis

Figure 3A shows the overlapping of Amide I absorption bands from 1590 to 1720 cm$^{-1}$ of Alpha, Gamma and Omicron SARS-CoV-2 variants. Slight differences can be recognized in the absorption band shape and spectral maxima. Looking at Fig 3A, a slight shift of Amide I absorption maxima occurs when comparing the spectra of Alpha, Gamma and Omicron S1 proteins. The maximum is located at (1650 ± 1) cm$^{-1}$ for Alpha Amide I absorption band, the same as for the WT S1 protein, as reported in the previous studies (58,59). In contrast, the Gamma and Omicron Amide I absorption bands have maxima located to (1648 ± 1) cm$^{-1}$ and (1647 ± 1) cm$^{-1}$, respectively.

These differences could be further highlighted by calculating the differences A(ω)(Gamma)-A(ω)(Alpha), and A(ω)(Omicron)-A(ω)(Alpha) (orange and green lines in Figure 3B, respectively) and comparing them, for instance, with the reproducibility of the absorption measurements (58,59), estimated by the difference between A(ω)(Alpha)-A(ω)(Alpha) for two different measurement runs (blue line in Figure 3B).

A sizable difference was observed well above the reproducibility of the absorption spectra measurements, which is approximately ±1%. In particular, both Gamma and Omicron SARS-CoV-2 S1 have a lower absorption intensity between 1665–1700 cm$^{-1}$ (in agreement with the differences observed in Figure 3A), and a slightly more intense signal at lower frequencies.

These spectral differences correspond to a different distribution of spectral absorption components, in terms of shifts in frequencies and/or variations in absorption intensity and therefore reflect different

IR vibrations of protein's intrinsic structures. To better understand this behavior, amide I absorption bands of S1 monomeric variant proteins were deconvoluted into Gaussian components using a multi-Gaussian fitting approach (as described in the section Materials and Methods). The results are displayed in Figure 4A, 4B and 4C for Alpha, Gamma and Omicron, respectively. In each panel, the convoluted bands are represented by underlying bars, whose height corresponds to the percentage contribution of each spectral component to the overall Amide I band intensity. The assignment to a specific secondary structure is indicated by different colors (orange for $\beta$-sheet, yellow for random coil, purple for $\alpha$-helix, green for $\beta$-turn and brown for side chain contribution). Table S1 in S4 summarizes Amide I vibrational frequencies for each S1 variant protein and their assignment to secondary structures. Notably, the Amide I band deconvolution of S1 protein from the WT has already been studied and reported in our previous works (58,59).

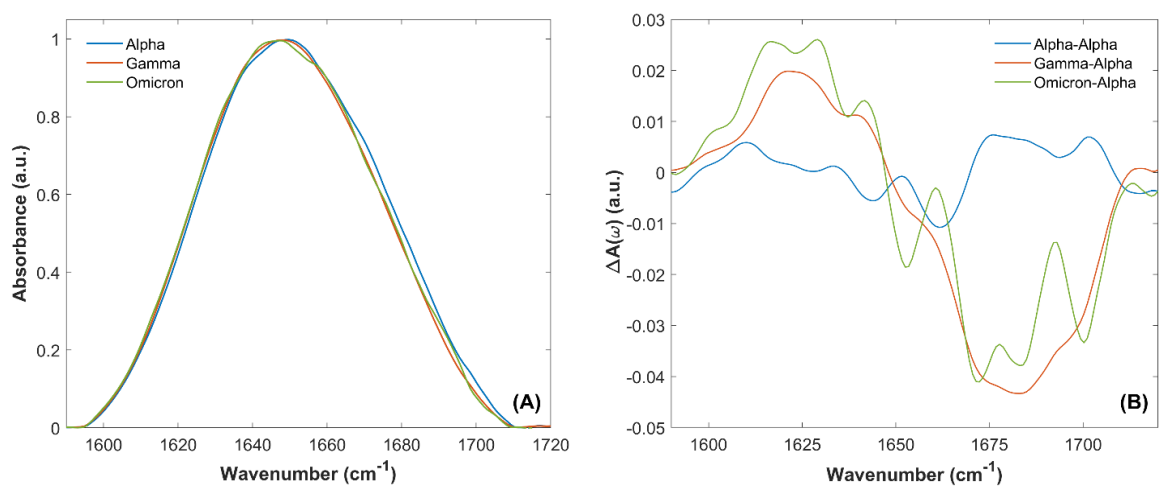

**Figure 3.** Comparison of amide I S1 absorption spectra for the three variants between 1590 and 1720 cm$^{-1}$. (A) Comparison of the S1 absorption spectra of Alpha (blue curve), Gamma (orange curve) and Omicron (green curve) variants of SARS-CoV-2 virus. (B) A($\omega$)(Gamma)-A($\omega$)(Alpha) (orange line) and A($\omega$)(Omicron)-A($\omega$)(Alpha) (green line) differences compared to the reproducibility of the SARS-CoV-2 S1 Alpha absorption spectrum. This reproducibility was estimated by the difference of A($\omega$)(Alpha)-A($\omega$)(Alpha) measured in two different measurement runs (blue line). A sizeable difference (well beyond the reproducibility of the individual absorption spectra) was observed when comparing the absorption of Gamma and Omicron with the Alpha one.

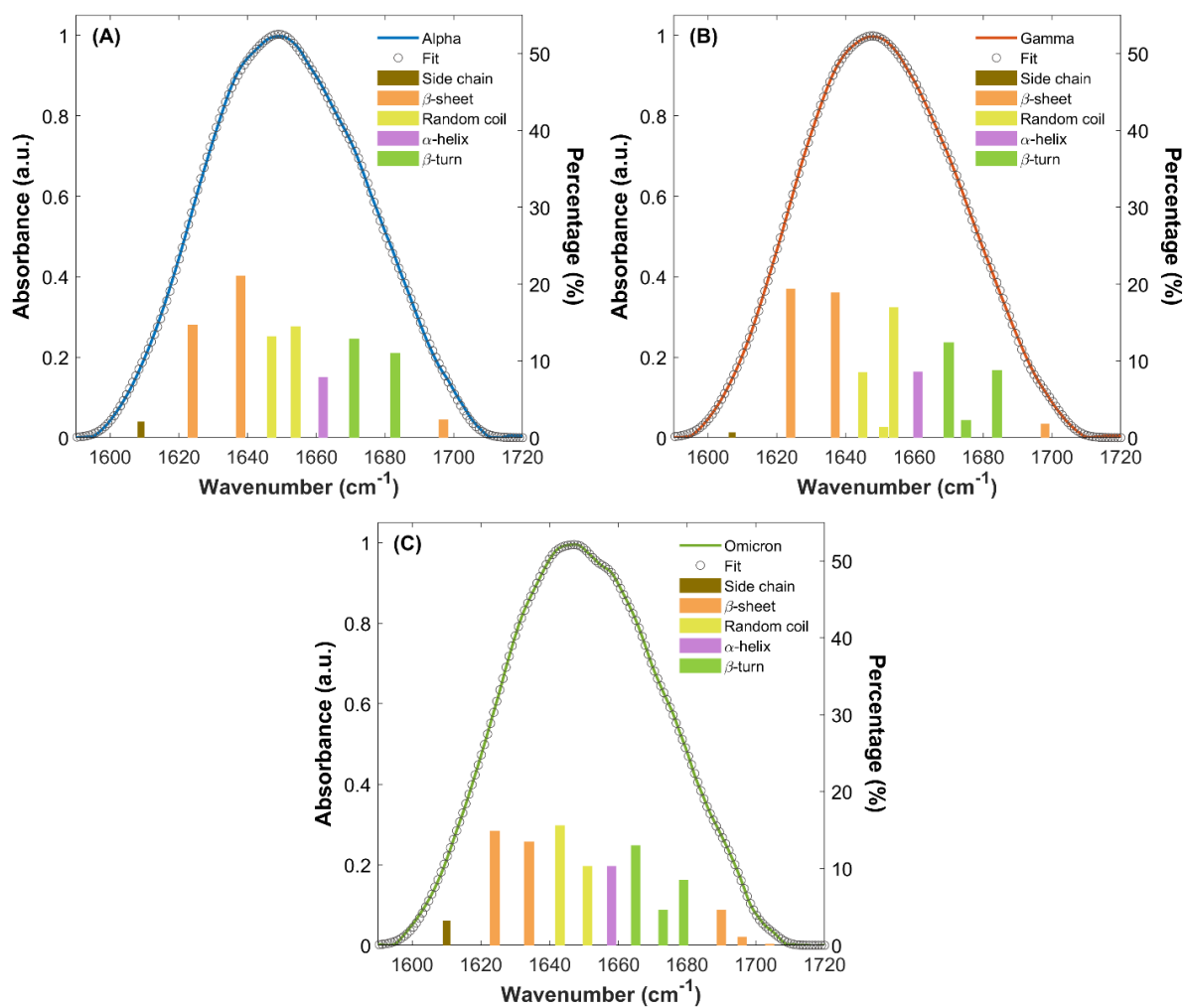

**Figure. 4.** Amide I absorption bands of S1 proteins for the three variants. IR amide I band of Alpha (A), Gamma (B) and Omicron (C) S1 proteins and their global fitting (grey circles curve), referring to the left y-axis. For each amide band, the deconvolution into Gaussian components is represented through the underlying bars, with assignments described through their colors (as explained in the legend). Their heights show the percentage contribution of each spectral component (right y-axis).

For all three S1 proteins, $\beta$-sheet vibrations (orange bars in Figure 4A, 4B and 4C) arise with two different contributions (see in S4 Table S1), one at low frequencies attributable to $\nu\perp$ mode, and the other at high frequency attributable to $\nu//$ mode (37-39, 58,59). This doublet absorption is generally associated to an antiparallel $\beta$-sheet structure (37-39, 58,59). In particular, both Alpha and Gamma S1 proteins show two low frequency $\beta$-sheet $\nu\perp$ absorptions around 1624 and 1637 cm$^{-1}$ and one high frequency $\beta$-sheet $\nu//$ absorption at about 1697 cm$^{-1}$. Omicron S1 protein exhibits two $\beta$-sheet $\nu\perp$ absorption peaks lying at 1624 and 1634 cm$^{-1}$, slightly lower in frequency compared to Alpha and Gamma proteins $\beta$-sheet $\nu\perp$ vibration modes and with an evident lower percentage contribution. At high frequency, Omicron amide I band shows three β-sheet ν// bands around 1690, 1696 and 1704 cm$^{-1}$, rather providing a greater contribution compared to Alpha and Gamma.

Both in Alpha, Gamma and Omicron S1 protein, α-helix vibrational mode (purple bar in Figure 4A, 4B and 4C) rises as a single absorption peak. It lies around 1661 cm$^{-1}$ for Alpha and Gamma S1 proteins, while it is slightly redshifted in Omicron S1 protein located at 1658 cm$^{-1}$.

Random coil (yellow bars in Figure 4A, 4B and 4C) structure vibrations are found between 1643 cm$^{-1}$ and 1654 cm$^{-1}$. In the Alpha protein amide I band, they rise at 1647 and 1654 cm$^{-1}$, while in Gamma protein they are found at 1645 and 1654 cm$^{-1}$, together with a weak shoulder at 1651 cm$^{-1}$. In Omicron S1 protein random coil vibrations are slightly redshifted, lying at 1643 and 1651 cm$^{-1}$.

Finally, β-turn (green bars in Figure 4A, 4B and 4C) contributions are found between 1665 cm$^{-1}$ and 1684 cm$^{-1}$. They rise as two absorption bands at 1671 and 1683 cm$^{-1}$ in Alpha S1 protein, while three β-turn peaks are found at 1670, 1675 and 1684 cm$^{-1}$ in Gamma amide I band. In Omicron protein amide I band, β-turn structures rise with three absorptions placed at 1665, 1673 and 1679 cm$^{-1}$, again shifted to lower frequencies compared to Alpha and Gamma amide band.

The percentage content of each secondary structure was estimated by calculating the ratio of the integrated intensity of its spectral components to the total integrated intensity of the amide I band (58-62), after the subtraction of side chain vibrations from the total area (59). Results are reported in Table 2 for each variant, expressed as the average and the standard deviation of the estimation of each secondary structure content across different protein depositions, as explained in the Materials and Methods section.

**Table 2.** Secondary structure percentage content from IR analysis. Percentage values were estimated from Amide I band deconvolution with a multi-Gaussian fit for monomeric S1 protein of SARS-CoV-2 Alpha, Gamma, and Omicron variants. Percentage data are presented as MEAN ± Standard deviation (SD).

|         | β-sheet (%) | α-helix (%) | β-turn (%) | Random coil (%) |
|---------|-------------|-------------|------------|-----------------|
| Alpha   | 39 ± 2      | 8 ± 1       | 25 ± 3     | 28 ± 2          |
| Gamma   | 40 ± 3      | 9 ± 1       | 24 ± 2     | 27 ± 1          |
| Omicron | 35 ± 3      | 11 ± 1      | 27 ± 1     | 27 ± 1          |

## 3.3 AlphaFold2 prediction and Molecular dynamics simulations

S1 proteins of Alpha, Gamma and Omicron variants were modelled using AlphaFold2, starting from their amino acid sequences (see S1), provided by Sino Biological Europe GmbH (Eschborn, Germany). According to AlphaFold2 predictions, S1 protein appears to assume two possible different configurations. In the first, the protein is stabilized in a "closed" state, with the RBD and NTD domains placed very close to each other and providing somehow a compact and folded conformation. In the second, S1 is stabilized in an "open" state, with the RBD and NTD domains placed further from each other if compared to the "closed" state (see S5, Figure S2). The distinction between the "open" and the "closed" state is based on the distance between the centers of mass of NTD and RBD, which are reported in Table S2, paragraph S5. It appears that, for each variant, the "closed" state has NTD-RBD distance lower than 5 nm, while the "open" configurations show NTD-RBD distance larger than 5 nm.

In order to predict and simulate how the three different proteins behave and adapt in aqueous environment at serological condition pH 7.4, MD simulations have been performed via GROMACSv. 2022.3 package (63,72) on Alpha, Gamma and Omicron protein models. To evaluate the goodness of the predictions, the Local Distance Difference Test (plDDT) was performed for all three variants, observing that either "open" and "closed" states present a similar plDDT value. Therefore, MD simulations were performed starting from both the "open" and the "closed" states for each variant. CHARMM22* force field was applied, and the proteins were let evolve for 600 ns, a time sufficient for a reasonable numerical convergence (64, 85). This was evaluated by observing the Root Mean Square Deviations (RMSD) values of atomic positions throughout the simulation, as reported in SI (see Table S3 and Figure S3). The time-dependent behavior of the Radius of Gyration (Rg) was studied for both closed and open models for all three S1 proteins (see S6, Figure S4). For Alpha, Gamma and Omicron, closed models retained a quite stable Rg value and preserved the closed configuration for the whole time. In the case of the initial open models of Alpha and Gamma, these tended to rearrange over time, decreasing their Rg values until reaching a closed state. On the other hand, the initial open model of Omicron maintained a stable Rg value throughout the simulation, remaining in an open state with a large Rg value. Initial and final Rg values for both "open" and "closed" states were calculated as the average over the first 50 ns and the last 100 ns of simulation, respectively. The results are reported in Table 3, while Rg curves for all six models are provided in SI (see S6, Figure S4).

**Table 3.** Radius of Gyration (Rg) values in nm for the three variants of S1 proteins. Results are reported for Alpha, Gamma and Omicron both in closed and open state. Values are obtained as the MEAN ± SD over the last 100 ns of MD simulations.

| Variants | Radius of Gyration (nm) | | | |
|---|---|---|---|---|
| | Closed state | | Open state | |
| | Initial | Final | Initial | Final |
| Alpha | 3.6 ± 0.1 | 3.33 ± 0.04 | 4.2 ± 0.3 | 3.5 ± 0.3 |
| Gamma | 3.20 ± 0.07 | 3.12 ± 0.02 | 3.5 ± 0.2 | 3.06 ± 0.04 |
| Omicron | 3.31 ± 0.05 | 3.18 ± 0.03 | 4.2 ± 0.2 | 4.3 ± 0.1 |

As expected, the final Rg values, calculated as the average over the last 100 ns of simulation, correspond to the most stable configuration, having the lowest value of Free Energy Surface (FES). FES values have been calculated and two-dimensional heatmaps were computed as a function of RMSD and Rg values. They are shown in Figure 5 for Alpha, Gamma and Omicron variants, respectively, for both initial "closed" (left panels) and "open" states (right panels). Generally, it can be observed that for all three variants, the systems derived from the initial "closed" (left panels in Figure 5) state appear to be more stable in comparison with the ones derived from the initial "open" state (right panels in Figure 5). In fact, in the first case the diffusion area of points in the phase space [Rg ; RMSD] is smaller with respect to the diffusion area in the second case, meaning that the protein needs to explore a small range of Rg values in order to achieve the equilibrium condition. In contrast, proteins derived from initial "open" state needs to span a larger space before finding the most stable configuration. In particular, Omicron S1 protein derived from the initial "open" state (Figure 5F) is the only one exploring a range of greater Rg, from 4.0 to 4.5 nm and here it finds an equilibrium condition, differently from the other protein models which assume only smaller Rg values throughout the whole simulations.

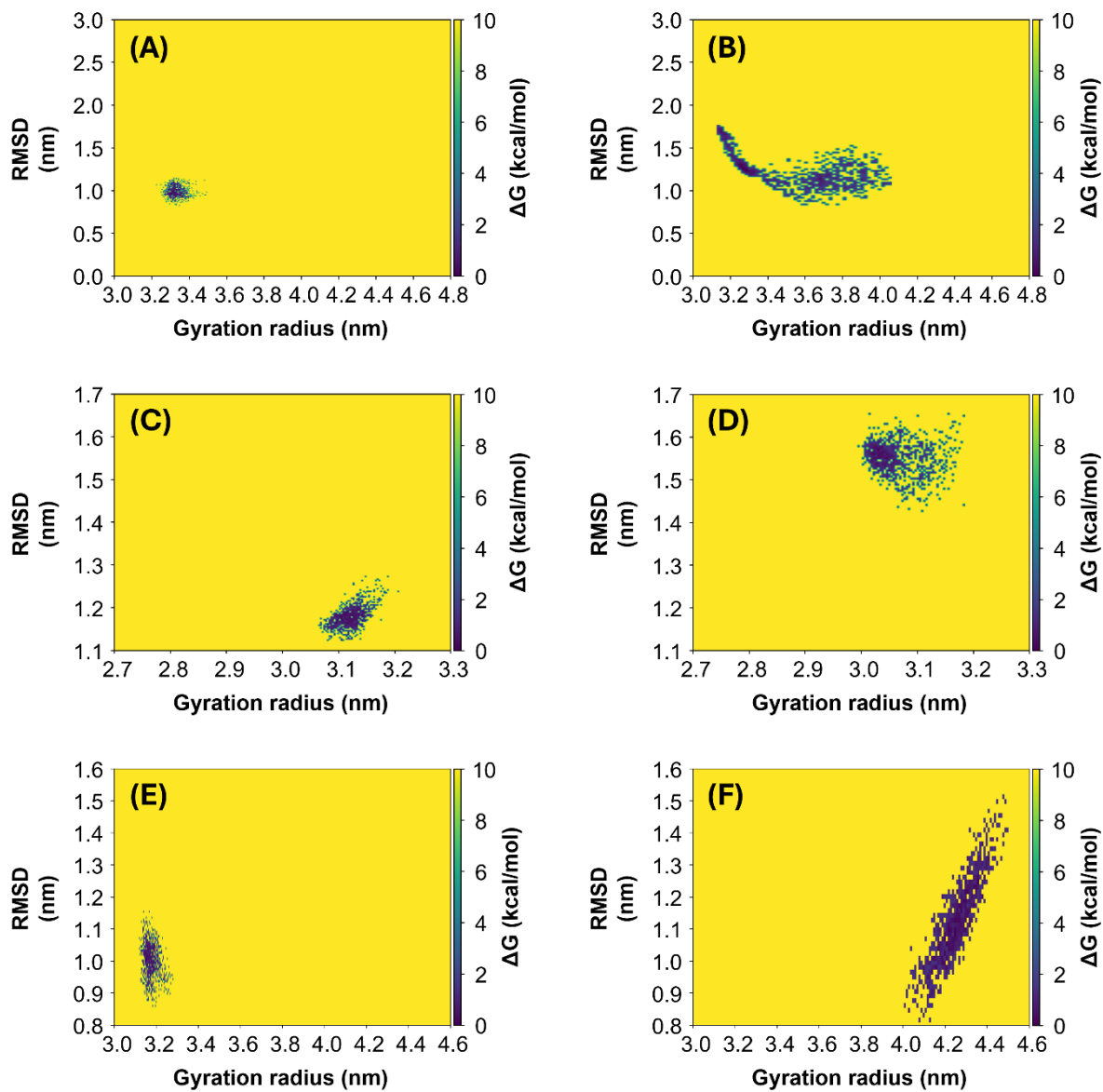

**Figure 5.** FES maps for the three variants of S1 protein. FES maps of Alpha (A-B), Gamma (C-D) and Omicron (E-F) are computed over the last 100 ns of each MD simulation, both for the systems starting from the initial "closed" state (left panels) and from the "open" state (right panels). The data are represented in the phase [Rg;RMSD], where each point represents the protein configuration recorded with a Dt=0.1 ns.

Computational results for the percentage contents of secondary structures are reported in Table 4. The results are calculated as the average and standard deviation of each secondary structure content over the last 100 ns of the simulation. Given the different behavior of Rg values for the three variants' S1 proteins, we report the percentage content of secondary structures in the "closed" states for Alpha and Gamma, and as the average between the "open" and "closed" models' values for Omicron.

**Table 4.** Secondary structure percentage content from MD analysis. Results for Alpha, Gamma and Omicron variants of SARS-CoV-2 virus obtained with MD simulations. Percentage values are calculated as the MEAN ± SD over the last 100 ns of simulation. For Omicron the results are obtained through an average between "open" and "closed" states.

|          | β-sheet (%) | α-helix (%) | β-turn (%) | Unordered (%) |
|----------|-------------|-------------|------------|---------------|
| Alpha    | 37 ± 3      | 6 ± 2       | 29 ± 2     | 28 ± 3        |
| Gamma    | 39 ± 2      | 7 ± 1       | 27 ± 2     | 27 ± 2        |
| Omicron  | 35 ± 2      | 6 ± 2       | 29 ± 2     | 30   3        |

### 3.4 Hydrophilic calculation and surface polarity computation

Hydrophilicity properties of the three VoCs S1 proteins were evaluated first computing their Gravy value using ProtParam web-server (https://web.expasy.org/protparam/). The closer Gravy value is to zero, the greater is the protein hydrophobicity. The results show that the Alpha and Omicron S1 proteins have the same Gravy value of -0.268, while Gamma S1 protein has a Gravy value of -0.246.

This kind of calculation depends solely on the hydropathy of the individual amino acids, and therefore only on the protein's primary structure. In order to verify how the 3D conformation of the three S1 proteins influences their hydrophilic properties and interaction with the solvent, Non-polar to polar (NPP) ratio surfaces (83,84) were computed using Protein-sol software (https://protein-sol.manchester.ac.uk/). The results are presented for S1 proteins of Alpha, Gamma and Omicron variants in the final configuration (600 ns), as provided by MD simulation. In particular, RBD and NTD are highlighted, ss these are of regions where the majority of mutations occur (see Figure 1). For the Alpha and Gamma variants, only the "closed" states surfaces are shown, as both their initial "open" and "closed" states reached a "closed" configuration by 600 ns. For Omicron S1 protein instead, both the "open" and "closed" states are shown. Noticeable differences can be observed in the NPP ratio behavior of RBDs. Alpha variant RBD shows a predominantly neutral/hydrophobic behavior (white/green region in Figure 6), in contrast to the Gamma and Omicron RBDs (both in "open" and "closed" states), which reveal strongly hydrophilic areas (purple regions Figure 6).

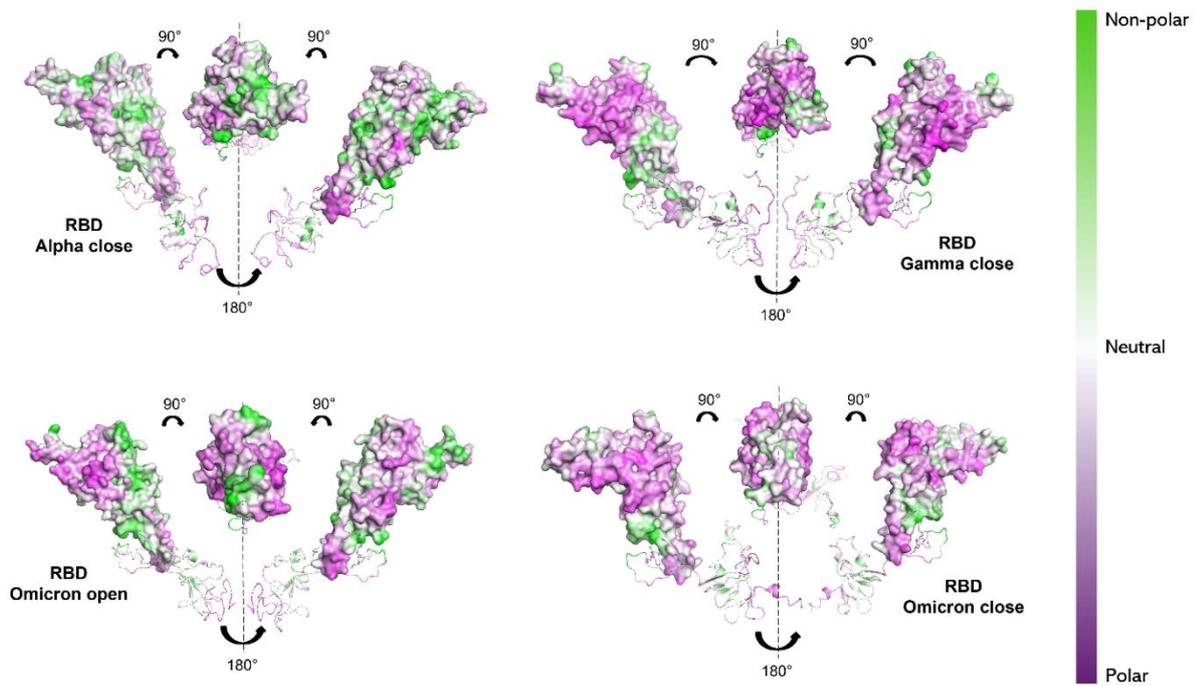

**Figure 6.** NPP surface ratio of the RBDs for the three variants. NPP ratio distribution is computed for RBDs surface for Alpha (closed state), Gamma (closed state) and Omicron (open and closed state). The color scale is shown on the right: green, white and purple color correspond to hydrophobic, neutral and hydrophilic behavior, respectively.

Observing the NTDs, Alpha variant still shows a more hydrophobic character (white/green regions in Figure 7) compared to the other two variants. In particular, Omicron (both in "open" and "closed" states) exhibits the most hydrophilic NTD character (purple regions in Figure 7), compared to Gamma and Alpha NTDs.

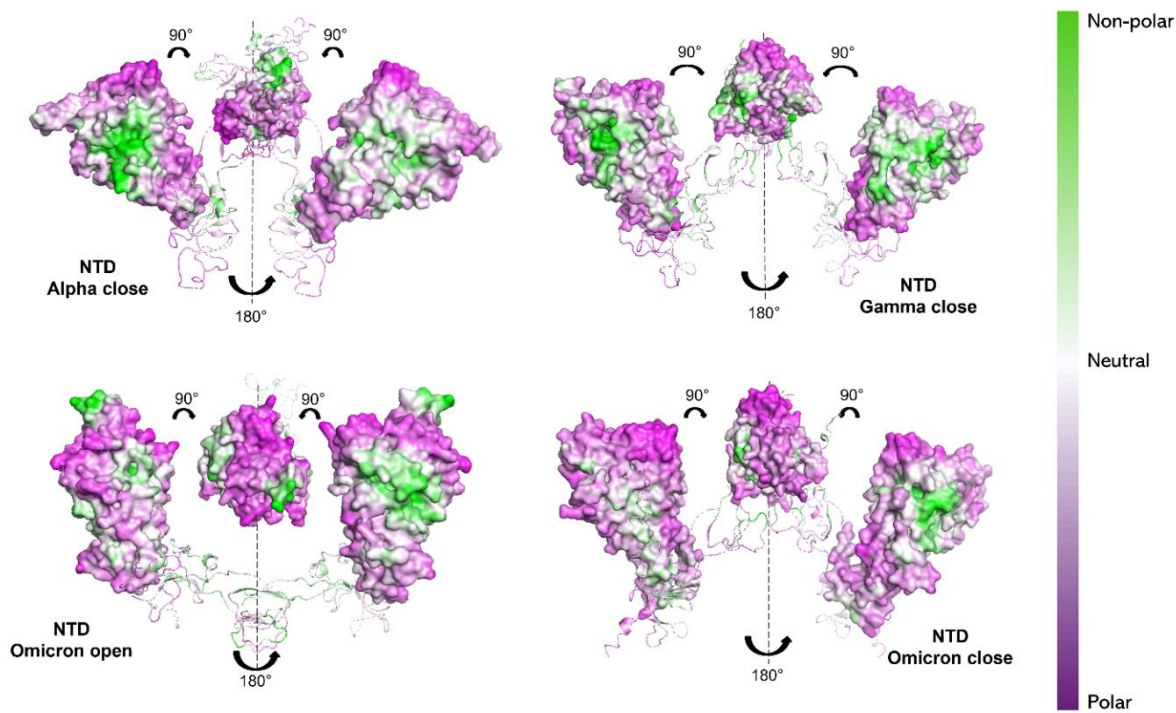

**Figure 7.** NPP surface ratio of the NTDs for the three variants. NPP ratio distribution is computed for NTDs surface for Alpha S1 protein (closed state), Gamma S1 protein (closed state) and Omicron S1 protein (open and closed state). The color scale is shown on the right: green, white and purple color correspond to hydrophobic, neutral and hydrophilic behavior, respectively.

## 4. Discussion

SARS-CoV-2 VoCs emerged as a result of multiple mutations in the genomic sequence of the SARS-CoV-2 virus during the pandemic evolution, leading to structural differences in both their spike and nucleocapsid proteins. It is known that these changes influence proteins behavior, especially in terms of interaction bridges and strength with ACE2 human receptor, binding affinity (16-19), flexibility (20,21), and accessibility (14,22). The study of mutations reveals that not all domains of the S protein are equally susceptible to mutations. S1 subunit, the one that actively anchors the ACE2 receptor through the RBD domain, harbors most mutations, compared to S2 subunit. In particular, NTD and RBD, located in S1 subunits, are the regions presenting the most important variations (14).

Here, we will discuss and compare experimental IR and CD data, as well as computational results, to detect differences induced by amino acid sequence mutations in the S1 subunit of Alpha, Gamma and Omicron VoCs of SARS-CoV-2 virus.

In particular, CD data suggest that mutations in amino acid sequences induced some structural changes in S1 subunits. These data have been deepened employing IR spectral data obtaining information on protein secondary structures, 3D conformation, dynamics, hydrophobicity and structural order. Initially, we investigated VoCs secondary structure content and provided their estimation combining the three techniques. Subsequently, a deep IR spectral analysis revealed other kinds of variations in VoCs proteins, in terms of hydrophobicity and 3D conformation, which have been further corroborated through the NPP surface computation and MD simulations.

The effects of VoCs mutations on S1 protein structure have been first observed in the CD spectra between 190-230 nm, i.e. in the region of the π-π * and n-π * electronic transitions of the peptide group (Figure 2). The evidence of structural changes results from recognizable variations at low wavelengths, where S1 from Gamma and Omicron variants exhibit a more intense and broader peak around 194 nm compared to the Alpha variant. This is due to the different content of CD-active amino acids in each subunit, such as Alanine, Valine, Leucine and Asparagine (86) and the competitive contribution of $\beta$-sheet and $\beta$-turn structures.

Significant differences in the CD spectra also occur between 200 and 240 nm: here, protein dichroism is generally very responsive to the abundance of $\alpha$-helix, $\beta$-sheet and coil structures (40,43,65). Changes in CD spectra recorded for S1 protein of three VoCs are the result of mutations accumulating in the VoCs amino acid sequences and of the subsequent adaptations of their secondary structure.

To further shed light on possible structural modifications occurring in S1 proteins due to their mutations, the IR amide I absorption band was inspected, revealing significant differences in the three variants (Figure 3 and Figure 4).

Indeed, IR deconvolution analysis highlights different shapes and distributions of amide I spectral components, due to different vibrations of proteins' structures. Among the most evident variations, $\beta$-sheet $\nu_\parallel$ modes occur as three distinct components in Omicron amide I band (see Figure 4C and see S4, Table S1) accounting for 6 % of the whole protein secondary structure. In contrast, amide I spectra of Alpha (Figure 4A) and Gamma (Figure 4B) variants show one single $\beta$-sheet $\nu_\parallel$ component, corresponding to about 2 % of the total intensity. On the other hand, in all three VoCs $\beta$-sheet $\nu_\perp$ vibrations occur with two intense peaks at low frequencies, but they contribute with a higher integrated intensity in Alpha and Gamma (among 36% and 38%), while they only constitute a lower percentage in Omicron absorption (around 28%).

Notable differences are also observed in the $\alpha$-helix absorption band, which results to be more intense in Omicron S1 protein (about 10 %), compared to Alpha and Gamma (about 8 %) and, finally, slight changes concern $\beta$-turn absorptions, occurring as two peaks in Alpha amide bands and as three distinct peaks Gamma and Omicron.

The influence of mutations on S protein secondary structure has been just slightly studied (17,20,26), mostly employing computational methods, which assume that mutations cause small changes in the 3D distribution, conformation and flexibility of protein structures.

In Tables 1, 2 and 4 we provide the secondary structures contents of each protein variant obtained through CD and IR spectroscopy data and MD simulations.

In Figure 8, we further provide a graphical comparison of the secondary structure fractions for the S1 protein of each VoC. Results are reported as the weighted average of the outcomes obtained for each type of structure and each VoC from CD, IR and MD approaches. Moreover, a more detailed comparison of results provided separately by the three different techniques, along with their potentialities and limitations are presented in SI, paragraph S7.

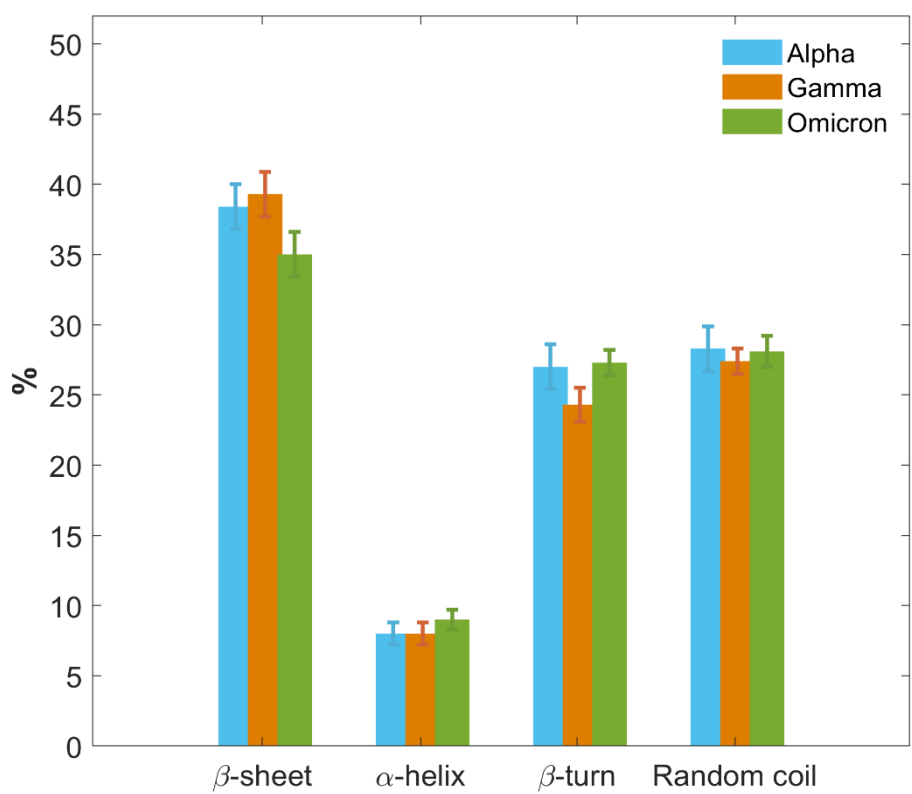

**Figure 8.** Histograms of secondary structure contents of S1 proteins for the three VoCs. Secondary structure percentage content S1 protein of Alpha (blue), Gamma variant (orange) and Omicron variant (green) are graphically and comparatively reported. Results are calculated as the average of CD, IR and MD data. Uncertainties are calculated through the standard deviation from average values.

The three VoCs show a similar secondary structure content, with some variations particularly concerning the Omicron S1 protein. All three proteins primarily consist of regions of disordered amino acids arrangements, with a random coil content around 28 % for all of them, while the β-turn percentage is around 27 % for Alpha and Omicron, and slightly lower in Gamma, around 24 %.

Meanwhile, all VoCs show a low content of α-helix structure, being the same for all three S1 proteins within the error, between 8% and 9%.

A noticeable difference can be observed in the β-sheet content. Alpha and Gamma S1 proteins show the same percentage content of β-sheet within the error bars, specifically between 38 % and 39 %. The Omicron S1 protein instead shows a lower percentage of β-sheet content, approximately 35 %, which represents a significant difference with respect to Alpha and Gamma beyond the error bars.

Moreover, amide I band maxima (see Figure 3) are located at 1650, 1648 and 1647 cm$^{-1}$ for the S1 proteins of Alpha, Gamma and Omicron variants, respectively. Note that the slight redshift of amide I from Alpha to Omicron seems to follow the increasing number of mutations, which is smaller for Alpha (seven) and maximum for Omicron (thirty-one) (see Figure 1).

Focusing more deeply on the deconvoluted IR spectral components (see Figure 4 and Table S1), the Omicron amide I clearly shows also a general slight shift to lower frequencies of all its components with respect to the other two protein variants.

The shift to lower frequencies in the amide I vibration can typically be linked to stronger hydrogen bonding interactions with surrounding water molecules. This increased hydrogen bonding results in a decrease in the strength of the C=O vibration, leading to redshift in the observed frequency (49,50, 59) (see SI, Fig. S6).

Our results would then suggest a more hydrophilic behavior of Omicron S1 protein (89,90). From NPP ratio patches of proteins' surfaces, Omicron RBD actually shows larger regions with lower NPP values, therefore larger hydrophilic area, if compared with Alpha RBD, which instead shows mostly neutral or hydrophobic regions (see Figure 6). Still, Omicron NTD shows wider hydrophilic areas if compared with both Alpha and Gamma NTDs (see Figure 7).

Gravy values are found to be very similar between the three variants S1 proteins (Alpha Gravy value=-0.268, Gamma Gravy value=-0.246 and Omicron Gravy value=-0.268), as expected from their high similarity values. Indeed, as determined using the Pairwise Sequence Alignment Emboss Needle (https://www.ebi.ac.uk/Tools/psa/emboss_needle/), Alpha and Gamma present a similarity of 98.4%, Alpha and Omicron about 96.8%, finally Gamma and Omicron of 95.6%. Protein alignments are reported in SI (see pdf files in SI). Nevertheless, Gravy values depend only on amino acids sequence, therefore the more hydrophilic behavior of Omicron S1 protein revealed from NPP surface calculation and IR measurement, suggests a difference in Omicron S1 protein 3D conformational structure, with respect to Alpha and Gamma S1.

The more hydrophilic character of Omicron S1 protein can also explain its lower content of β-sheet structure with respect to Alpha and Gamma variants (see Figure 8). Indeed, the formation of β-sheet structures is known to be favored by hydrophobic amino acids, and they are also usually associated with a higher average hydrophobic value (89,90). Moreover, due to the hydrophobic general character of proteins structure, β-sheets usually tend to form in the buried region of the protein (91). Therefore, the decreasing of β-sheets in Omicron S1 could also be related to its tendency to assume an "open" conformation, more exposed to the surrounding solvent, as found out through MD simulations. Indeed, computational results have recognized an important variation in the three VoCs Rg values (see Table 3). Alpha and Gamma S1 proteins, although starting from an "open" configuration, tend to gradually fold and close on their structure, decreasing their Rg values until a "closed" state is achieved with a final Rg value of (3.5 ± 0.3) nm and (3.06 ± 0.04) nm, respectively. Omicron, instead, is the only VoC which does not decrease its Rg value, rather it retains an "open" configuration if starting from an "open" model, with a final Rg value of (4.3 ± 0.1) nm (92).

This MD outcome suggests that Omicron S1 protein appears more inclined to preserve its "open" configuration, where RBD and NTD domains are further apart than in the "closed" configuration (see Figure S2).

Assuming that Omicron S1 protein can on average assume equally "open" and a "closed" state in water, while Alpha and Gamma S1 protein generally assume only a "closed" state, this conformational difference due to the increasing number of mutations, can find confirmation in the different content of β-sheet secondary structure and in the spectral difference observed in IR and CD spectra.

The open and closed configurations can be confidently identified as the "up" and "down" configurations of the RBD in the entire S protein, respectively. (93-96). Therefore, our results could suggest Omicron S protein has a stronger stability in the RBD-up configuration compared with Alpha and Gamma. Considering the infectious process, it is clear from the literature that S protein assumes the RBDs in the "up" conformation in order to bind ACE2 on the surface of target cells before undergoing viral uptake and fusion (97-99). Our experimental and computational results would be in agreement with the easier trend of Omicron to bind the receptor and subsequently, with its higher level of affinity and infectivity with respect to other variants (16,19,22,30). Actually, as shown in literature (100) mutations characterizing the S protein of Omicron variant, and in particular its RBD, seem to generally facilitate a more efficient RBD "down" to "up" conformation and the attachment of ACE2, in accordance with our results. More specifically, S1 of Omicron seems to prefer a one-

RBD-up conformation (16,101), which nevertheless provides a stronger interaction with ACE2 and an enhanced and prolonged viral attachment (16). Still, the more hydrophilic behavior characterizing Omicron S1 protein and recognized from IR results and NPP ratio surface also contributes in principle to the stronger capability of this VoC to bind the ACE2 receptor. Some studies have recognized the importance of hydration forces in complex stability, dynamic and binding kinetics, making the binding site accessible to both ligand and receptor (87).

## 5. Conclusions

In this paper, we perform for the first time, to the best of our knowledge, a systematic and comparative investigation of monomeric S1 proteins of Alpha, Gamma and Omicron variants of SARS-CoV-2 virus. We focused on the S1 protein domains of these VoCs which are responsible for the anchoring to the host receptor ACE2, at serological pH (7.4), revealing differences induced by the mutations in amino acids sequences and interpreting the results in terms of their secondary structure, hydrophobicity and conformational structure.

An extensive survey and description of the structural features of S1 proteins is provided, employing both experimental approaches and computational methods. The evident changes in their CD spectra proved that the three mutants have significant differences in their structure.

Combining the results from IR amide I band (1600-1710 $cm^{-1}$) and dichroic signal (190-240 nm) analysis together with results from MD simulations, the secondary structure content has been estimated (in terms of $β$-sheet, random coil, $α$-helix and $β$-turn contents) for the three VoCs S1 proteins. Exploiting all three techniques, it was shown that the Alpha, Gamma and Omicron S1 proteins have very similar secondary structure content, as expected from the high similarity of their amino acid sequences and the few amino acid mutations, nevertheless showing not negligible differences, out of the error, concerning the lower $β$-sheet content of Omicron S1 compared to Alpha and Gamma.

Moreover, the redshift of Omicron amide I band components with respect to those of Alpha and Gamma, could be associated with its higher hydrophilic character. This has been further testified by NPP ratio surface computed on RBDs and NTDs domains of the three S1 proteins, i.e. the regions where most mutations occur. Actually, S1 from the Omicron variant presents an overall hydrophilic character, showing larger high-NPP-ratio areas. This also agrees with its lower $β$-sheet content, which are usually associated to more hydrophobic environment and tend to form in buried regions of the protein. These results further suggest a variation of Omicron 3D-conformation which is then coherently testified by computational results. ColabFOLD predicted both "open" and "close" models for both Alpha, Gamma and Omicron S1 proteins and, performing MD simulation on both initial states, Omicron S1 protein results to be the only one retaining an "open" state when starting from an "open" state, while Alpha and Gamma "open" states tend to assume a "close" configuration after 600 ns. This can both be brought back to a 3D spatial difference concerning Omicron variant, partially justifying spectral variations, and could also in principle be associated to a stronger tendence of Omicron S1 protein to retain an "open" configuration, which coincides with the "up" state of S protein (RBD up). This is the configuration that S protein assumes when it anchors the ACE2 receptor, therefore the stronger tendency of Omicron S1 protein to retain an "open" state could be related to its stronger affinity with ACE2, compared to other variants. Moreover, in accordance with the hydrophilic behavior, mutations of Omicron variant have also been recognized to cause a remarkable change in its S protein electrostatic potential surface, which is much more positive compared to previous VoCs (19). This constitutes an important evolutionary adaptation of the virus ensuring a strongly enhanced interaction with the negatively charged surface of ACE2 receptor.

It is clear that the balance between hydrophilicity and hydrophobicity in the binding regions plays a primary role in the recognition and interaction between two proteins. Therefore, a quantification of both the hydrophilic and hydrophobic contribution, and therefore of the entire hydropathy profile, is useful to better characterize the interaction between two proteins. Moreover, it is known that hydrophobicity/hydrophilicity distribution of viral proteins plays a role in the viral transmissibility (102-104), affecting the exposed surface area and the protein stability, especially if dealing with viruses which are transmitted through air (therefore in a humid environment) (105,106).

In conclusion, our results combine different techniques, both experimental and computational, for the understanding of conformational structure of protein domains with high amino acid similarity, and for the research of structural changes induced by very few mutated amino acids. They confirm the excellent capability of IR and CD spectroscopies to provide rapid and insightful information on protein secondary structures, shedding light on various aspects, such as the hydrophobicity, the conformational order and functionalities, from each protein domain to complex S1 structure. The knowledge of the structural characteristics of SARS-CoV-2 S1 variant proteins represent a crucial step in the development of effective therapeutic protocols and/or prophylaxis and monitoring strategies and it is of primary importance for understanding and addressing further surveillance, preventive and monitoring actions.

**Supporting Information**

Supporting Information is available from the Wiley Online Library or from the author.

**Acknowledgements**

This work was funded by the NATO Science for Peace and Security Program under grant No. 5889 "SARS-CoV-2 Multi-Messenger Monitoring for Occupational Health & Safety (SARS 3M)", by European Union under Next Generation EU (NGEU) PRIN 2022 PNRR partnership "P2022NMBAJ" – "Ultrasensitive detEction oF vocs and pAthogens" and finally by the "Progetti di Ricerca (Piccoli, Medi) di Ateneo 2023" grant entitled "URGENt - UltRasensitive multi-messenGEr bioseNsing for bioaerosol". Financial support by the Grant to Department of Science, Roma Tre University (MIUR-Italy Dipartimenti di Eccellenza, ARTICOLO 1, COMMI 314-337 LEGGE 232/2016) is gratefully acknowledged. We acknowledge the CINECA award under the ISCRA initiative, for the availability of high-performance computing resources and support.

**Author contributions:** Conceptualization: A.D., T.M., S.L.; Methodology: A.D., T.M., N.L., V.M., A.N. and S.L.; Investigation: A.D., T.M., R.M. and S.M.; IR spectroscopy: A.D., T.M., R.M., and S.M. CD spectroscopy: T.M., R.M. and A.N. MD simulation: N.L. and V.M. Formal Analysis: A.D., T.M., N.L. and R.M.; Data interpretation: A.D., S.L., T.M.; Visualization: A.D., N.L., T.M., R.M. and S.M.; Supervision: S.L., A.D., V.M.; Funding acquisition: S.L., A.N., V.M., A.D. and T.M.; Writing—original draft: A.D., T.M., R.M. S.L.; Writing—review and editing: all authors. All authors have read and agreed to the published version of the manuscript.

**Conflict of interest statement**

The authors declare no conflict of interest.

**References**

[1] https://data.who.int/dashboards/covid19/deaths?n=c

[2] Cucinotta, D.; Vanelli, M. WHO Declares COVID-19 a Pandemic. Acta Biomed. 2020, 91, 157–160.

[3] M.M. Hatmal, W. Alshaer, M.A.I. Al-Hatamleh, M. Hatmal, O. Smadi, M.O. Taha, A.J. Oweida, J.C. Boer, R. Mohamud, M. Plebanski. Comprehensive Structural and Molecular Comparison of Spike Proteins of SARS-CoV-2, SARS-CoV and MERS-CoV, and Their Interactions with ACE2. Cells 9, 2638 (2020).

[4] P. Zhao, J. L. Praissman, O. C. Grant, Y. Cai, T. Xiao, K. E. Rosenbalm, K. Aoki, B. P. Kellman, R. Bridger, D. H. Barouch, M. A. Brindley, N. E. Lewis, M. Tiemeyer, B. Chen, R.

J. Woods, L. Wells, Virus-Receptor Interactions of Glycosylated SARS-CoV-2 Spike and Human ACE2 Receptor. Cell Host Microbe 28, 1–16 (2020).

[5] R. Yadav, J.K. Chaudhary, N. Jain, P.K. Chaudhary, S. Khanra, P. Dhamija, A. Sharma, A. Kumar, S. Handu. Role of Structural and Non-Structural Proteins and Therapeutic Targets of SARS-CoV-2 for COVID-19. Cells 10, 821 (2021).

[6] M.Z. Salleh, J.P. Derrick, Z.Z. Deris. Structural Evaluation of the Spike Glycoprotein Variants on SARS-CoV-2 Transmission and Immune Evasion. Int. J. Mol. Sci. 22, 7425 (2021).

[7] S. Kang, M. Yang, Z. Hong, L. Zhang, Z. Huang, X. Chen, S. He, Z. Zhou, Z. Zhou, Q. Chen, Y. Yan, C. Zhang, H. Shan, S. Chen. Crystal structure of SARS-CoV-2 nucleocapsid protein RNA binding domain reveals potential unique drug targeting sites. Acta Pharmaceutica Sinica B 10, 1228 (2020).

[8] Y. Peng, N. Du, Y. Lei, S. Dorje, J. Qi, T. Luo, G.F. Gao, H. Song. Structures of the SARS‐CoV‐2 nucleocapsid and their perspectives for drug design. EMBO J. 39, e105938 (2020)

[9] W. Yan, Y. Zheng, X. Zeng, B. He, W. Cheng. Structural biology of SARS-CoV-2: open the door for novel therapies. Sig. Transduct. Target. Ther. 7, 26 (2022).

[10] G. McLean, J. Kamil, B. Lee, P. Moore, T.F. Schulz, A. Muik, U. Sahin, Ö. Türeci, S. Pather. The impact of evolving SARS-CoV-2 mutations and variants on COVID-19 vaccines. MBio 13, e0297921 (2022).

[11] M. Sironi, S.E. Hasnain, B. Rosenthal, T. Phan, F. Luciani, M.A. Shaw, M.A. Sallum, M.E. Mirhashemi, S. Morand S, F. González-Candelas. SARS-CoV-2 and COVID-19: A genetic, epidemiological, and evolutionary perspective. Infect. Genet. Evol. 84, 104384 (2020).

[12] W.T. Harvey, A.M. Carabelli, B. Jackson, R.K. Gupta, E.C. Thomson, E.M. Harrison, C. Ludden, R. Reeve, A. Rambaut. SARS-CoV-2 variants, spike mutations and immune escape. Nat. Rev. Microbiol. 19, 409–424 (2021).

[13] D. Zella, M. Giovanetti, F. Benedetti, F. Unali, S. Spoto, M. Guarino, S. Angeletti, M. Ciccozzi.The variants question: What is the problem? J. Med. Virol. 93, 6479–6485 (2021).

[14] S. Fatihi, S. Rathore, A.K. Pathak, D. Gahlot, M. Mukerji, N. Jatana, L. Thukral. A rigorous framework for detecting SARS-CoV-2 spike protein mutational ensemble from genomic and structural features. Current Research in Structural Biology 3, 290 (2021).

[15] B. Imbiakha, S. Ezzatpour, D. W. Buchholz, J. Sahler, C. Ye, X.A. Olarte-Castillo, A. Zou, C. Kwas, K. O'Hare, A. Choi, R.A. Adeleke, S. Khomandiak, L. Goodman, M.C. Jager, G.R. Whittaker, L. Martinez-Sobrido, A. August, H.C. Aguilar. Age-dependent acquisition of pathogenicity by SARSCoV-2 Omicron BA.5., Sci. Adv. 9, eadj1736 (2023).

[16] R. Zhu, D. Canena, M. Sikora, M. Klausberger, H. Seferovic, A. R. Mehdipour, L. Hain, E. Laurent, V. Monteil, G. Wirnsberger, R. Wieneke, R. Tampé, N.F. Kienzl, L. Mach, A. Mirazimi, Y.J. Oh, J.M. Penninger, G. Hummer, P. Hinterdorfer. Force-tuned avidity of spike variant-ACE2 interactions viewed on the single-molecule level. Nature Communications 13, 7926 (2022).

[17] A.M. Carabelli, T.P. Peacock, L.G. Thorne, et al. SARS-CoV-2 variant biology: immune escape, transmission and fitness. Nat. Rev. Microbiol. 21, 162–177 (2023).

[18] C. Laffeber, K. de Koning, R. Kanaar, J.H.G. Lebbink. Experimental Evidence for Enhanced Receptor Binding by Rapidly Spreading SARS-CoV-2 Variants. J. Mol. Biol. 433, 167058 (2021).

[19] H.H. Gan, J. Zinno, F. Piano, K.C. Gunsalus. Omicron Spike Protein Has a Positive Electrostatic Surface That Promotes ACE2 Recognition and Antibody Escape. Front.Virol. 2, 894531 (2022).

[20] G. B. Chand, A. Banerjee, G. K. Azad. Identification of twenty-five mutations in surface glycoprotein (Spike) of SARS-CoV-2 among Indian isolates and their impact on protein dynamics. Gene Reports 21, 100891 (2020).

[21] S. Ozono, Y. Zhang, H. Ode, K. Sano, T. S. Tan, K. Imai, K. Miyoshi, S. Kishigami, T. Ueno, Y. Iwatani, T. Suzuki, K. Tokunaga. SARS-CoV-2 D614G spike mutation increases entry efficiency with enhanced ACE2-binding affinity. Nature Communications 12, 848 (2021).

[22] T.T. Nguyen, P.N. Pathirana, T. Nguyen, Q.V.H. Nguyen, A. Bhatti, D.C. Nguyen, D.T. Nguyen, N.D. Nguyen, D. Creighton, M. Abdelrazek. Genomic mutations and changes in protein secondary structure and solvent accessibility of SARS-CoV-2 (COVID-19 virus). Sci. Rep. 11, 3487 (2021).

[23] P.R.S. Sanches, I. Charlie-Silva, H.L.B. Braz, C. Bittar, M.F. Calmon, P. Rahal, E.M. Cilli. Recent advances in SARS-CoV-2 Spike protein and RBD mutations comparison between new variants Alpha (B.1.1.7, United Kingdom), Beta (B.1.351, South Africa), Gamma (P.1, Brazil) and Delta (B.1.617.2, India). Journal of Virus Eradication 7, 100054 (2021).

[24] M. Suleman, Q. Yousafi, J. Ali, S.S. Ali, Z. Hussain, S. Ali, M. Waseem, A. Iqbal, S. Ahmad, A. Khan, Y. Wang, D.Q. Wei. Bioinformatics analysis of the differences in the binding profile of the wild-type and mutants of the SARS-CoV-2 spike protein variants with the ACE2 receptor. Computers in Biology and Medicine 138, 104936 (2021).

[25] L. Wang, G. Cheng. Sequence analysis of the emerging SARS-CoV-2 variant Omicron in South Africa. J. Med. Virol. 94, 1728 (2022).

[26] V. Tragni, F. Preziusi, L. Laera, A. Onofrio, I. Mercurio, S. Todisco, M. Volpicella, A. De Grassi, C.L. Pierri. Modeling SARS-CoV-2 spike/ACE2 protein–protein interactions for predicting the binding affinity of new spike variants for ACE2, and novel ACE2 structurally related human protein targets, for COVID-19 handling in the 3PM context. EPMA Journal 13, 149–175 (2022).

[27] M.I. Barton, S.A. MacGowan, M.A. Kutuzov, O. Dushek, G. J. Barton, P. A. van der Merwe. Effects of common mutations in the SARS-CoV-2 Spike RBD and its ligand, the human ACE2 receptor on binding affinity and kinetics. eLife 10, e70658 (2021).

[28] I. Mercurio, V. Tragni, F. Busto, A. De Grassi, C. L. Pierri. Protein structure analysis of the interactions between SARS-CoV-2 spike protein and the human ACE2 receptor: from conformational changes to novel neutralizing antibodies. Cell. Mol. Life Sci. 78, 1501–1522 (2021).

[29] T. A. Shishir, T. Jannat, I.B. Naser. An in-silico study of the mutation-associated effects on the spike protein of SARS-CoV-2, Omicron variant. PLOS ONE 17, e0266844 (2022).

[30] J.T. Ortega, F.H. Pujol, B. Jastrzebska, H.R. Rangel, Mutations in the SARS-CoV-2 spike protein modulate the virus affinity to the human ACE2 receptor, an in silico analysis. EXCLI J. 20, 585-600 (2021).

[31] F. Piccirilli, F. Tardani, A. D'Arco, G. Birarda, L. Vaccari, S. Sennato, S. Casciardi, S. Lupi. Infrared Nanospectroscopy Reveals DNA Structural Modifications upon Immobilization onto Clay Nanotubes. Nanomaterials 11, 1103 (2021).

[32] G. Bálint, B. Vörös-Horváth, A. Széchenyi. Omicron: increased transmissibility and decreased pathogenicity. Sig. Transduct. Target. Ther. 7, 151 (2022).

[33] Y. Araf, F. Akter, Y.-D. Tang, R. Fatemi, Md. S. A. Parvez, C. Zheng, Md. G. Hossain. Omicron variant of SARS‐CoV‐2: Genomics, transmissibility, and responses to current COVID‐19 vaccines. J. Med. Virol. 94, 1825‐1832 (2022).

[34] I. Torjesen. Covid-19: Omicron may be more transmissible than other variants and partly resistant to existing vaccines, scientists fear. BMJ 375, 2943 (2021).

[35] E. Volz, S. Mishra, M. Chand, et al. Assessing transmissibility of SARS-CoV-2 lineage B.1.1.7 in England. Nature 593, 266–269 (2021).

[36] R. Dong, T. Hu, Y. Zhang, Y. Li, X.-H- Zhou. Assessing the Transmissibility of the New SARS-CoV-2 Variants: From Delta to Omicron. Vaccines 10, 496 (2022).

[37] A. Barth. The infrared absorption of amino acid side chains. Prog. Biophys. Mol. Biol. 74, 141–173 (2000).

[38] A. Barth. Infrared spectroscopy of proteins. Biochim. Biophys. Acta 1767, 1073–1101 (2007).

[39]  H. Yang, S. Yang, J. Kong, A. Dong, S. Yu. Obtaining information about protein secondary structures in aqueous solution using Fourier transform IR spectroscopy. Nat. Protoc. 10, 382–396 (2015).

[40]  S.M. Kelly, T.J. Jess, N.C. Price. How to study proteins by circular dichroism. Biochim. Biophys. Acta, Proteins Proteomics 1751, 119-139 (2005).

[41]  A.J. Miles, R.W. Janes, B.A. Wallace. Tools and methods for circular dichroism spectroscopy of proteins: a tutorial review. Chem. Soc. Rev. 50, 8400-8413 (2021).

[42]  A.J. Miles, B.A. Wallace. Circular dichroism spectroscopy of membrane proteins. Chem. Soc. Rev. 45, 4859-4872 (2016).

[43]  L. Whitmore, BA. Wallace. Protein secondary structure analyses from circular dichroism spectroscopy: methods and reference databases. Biopolymers. 89(5), 392-400 (2008).

[44]  C.H. Li, X. Nguyen, L. Narhi, L. Chemmalil, E. Towers, S. Muzammil, J. Gabrielson, Y. Jiang. Applications of circular dichroism (CD) for structural analysis of proteins: qualification of near‐ and far‐UV CD for protein higher order structural analysis. Journal of Pharmaceutical Sciences 100(11), 4642-4654 (2011).

[45]  U.J. Meierhenrich, J.J. Filippi, C. Meinert, J.H. Bredehöft, J.I. Takahashi, L. Nahon, N.C. Jones, S.V. Hoffmann. Circular Dichroism of Amino Acids in the Vacuum-Ultraviolet Region. Angewandte Chemie International Edition 49, 7799-7802 (2010).

[46]  S.D. Moran, M.T. Zanni. How to get insight into amyloid structure and formation from infrared spectroscopy. The journal of physical chemistry letters 5, 1984-1993 (2014).

[47]  F. Vosough, A. Barth. Characterization of homogeneous and heterogeneous amyloid-β42 oligomer preparations with biochemical methods and infrared spectroscopy reveals a correlation between infrared spectrum and oligomer size. ACS Chemical Neuroscience 12(3), 473-488 (2021).

[48]  S. Woutersen, P. Hamm. Time-resolved two-dimensional vibrational spectroscopy of a short α-helix in water. The Journal of Chemical Physics 115(16), 7737-7743 (2001).

[49]  E.S. Manas, Z. Getahun, W.W. Wright, W.F. DeGrado, J.M. Vanderkooi. Infrared spectra of amide groups in α-helical proteins: evidence for hydrogen bonding between helices and water. Journal of the American Chemical Society 122(41), 9883-9890 (2000).

[50]  N.S. Myshakina, Z. Ahmed, S.A. Asher. Dependence of amide vibrations on hydrogen bonding. The Journal of Physical Chemistry B 112(38), 11873-11877 (2008).

[51]  S.S.T. Gamage, T.N. Pahattuge, H. Wijerathne, et al. Microfluidic affinity selection of active SARS-CoV-2 virus particles. Sci. Adv. 8, eabn9665 (2022).

[52] R. Funari, K.Y. Chu, A.Q. Shen. Detection of antibodies against SARS-CoV-2 spike protein by gold nanospikes in an opto-microfluidic chip. Biosens Bioelectron. 169, 112578 (2020).

[53] B. Tan, X. Zhang, A. Ansari, et al. Design of a SARS-CoV-2 papain-like protease inhibitor with antiviral efficacy in a mouse model. Science 383, 1434–1440 (2024).

[54] N. Kumar, N.P. Shetti, S. Jagannath, T. M. Aminabhavi. Electrochemical sensors for the detection of SARS-CoV-2 virus. Chem. Eng. J. 430, 132966 (2022).

[55] J.H. Lee, J.W. Kim, H.E. Lee, et al. A dual-targeting approach using a human bispecific antibody against the receptor-binding domain of the Middle East Respiratory Syndrome Coronavirus. Virus Research 345, 199383 (2024).

[56] D.L. Kitane, S. Loukman, N. Marchoudi, A. Fernandez-Galiana, F.Z. El Ansari, F. Jouali, J. Badir, J.L. Gala, D. Bertsimas, N. Azami, O. Lakbita, O. Moudam, R. Benhida, J. Fekkak. A simple and fast spectroscopy-based technique for Covid-19 diagnosis. Sci Rep 11, 16740 (2021).

[57] M. Mirdita, K. Schütze, Y. Moriwaki, L. Heo, S. Ovchinnikov, M. Steinegger. ColabFold: making protein folding accessible to all. Nat Methods 19, 679–682 (2022).

[58] A. D'Arco, M. Di Fabrizio, T. Mancini, R. Mosetti, S. Macis, G. Tranfo, G. Della Ventura, A. Marcelli, M. Petrarca, S. Lupi. Secondary Structures of MERS-CoV, SARS-CoV, and SARS-CoV-2 Spike Proteins Revealed by Infrared Vibrational Spectroscopy. Int. J. Mol. Sci. 24, 9550 (2023).

[59] T. Mancini, S. Macis, R. Mosetti, N. Luchetti, V. Minicozzi, A. Notargiacomo, M. Pea, A. Marcelli, G. D. Ventura, S. Lupi, A. D'Arco. Infrared Spectroscopy of SARS-CoV-2 Viral Protein: from Receptor Binding Domain to Spike Protein. Adv. Sci. 2400823 (2024).

[60] M. Wolpert, P. Hellwig. Infrared spectra and molar absorption coefficients of the 20 alpha amino acids in aqueous solutions in the spectral range from 1800 to 500 cm− 1. Spectrochimica Acta Part A: Molecular and Biomolecular Spectroscopy 64(4), 987-1001 (2006).

[61] V. Ricciardi, M. Portaccio, G. Perna, M. Lasalvia, V. Capozzi, F.P. Cammarata, P. Pisciotta, G. Petringa, I. Delfino, L. Manti, et al. FT-IR Transflection Micro-Spectroscopy Study on Normal Human Breast Cells after Exposure to a Proton Beam. Appl. Sci. 11(2), 540 (2021).

[62] I. Delfino, M. Portaccio, B. Della Ventura, D.G. Mita, M. Lepore. Enzyme distribution and secondary structure of sol-gel immobilized glucose oxidase by micro-attenuated total reflection FT-IR spectroscopy. Mater. Sci. Eng. C 33, 304–310 (2013).

[63] D. Van Der Spoel, E. Lindahl, B. Hess, G. Groenhof, A.E. Mark, H.J.C. Berendsen. GROMACS: Fast, flexible, and free. J. Comput. Chem. 26(16), 1701–18 (2005).

[64] B. A. Russell, K. Kubiak-Ossowska, P. A. Mulheran, D. J. S. Birch and Y. Chen. Locating the nucleation sites for protein encapsulated gold nanoclusters: a molecular dynamics and fluorescence study. Phys. Chem. Chem. Phys. 17, 21935-21941 (2015).

[65] N. Sreerama, R.W. Woody. On the analysis of membrane protein circular dichroism spectra. Protein Science 13, 100-112 (2004).

[66] N. Sreerama, R.W. Woody. Estimation of Protein Secondary Structure from Circular Dichroism Spectra: Comparison of CONTIN, SELCON, and CDSSTR Methods with an Expanded Reference Set. Analytical Biochemistry 287, 252-260 (2000).

[67] J. Jumper, R. Evans, A. Pritzel, T. Green, M. Figurnov, O. Ronneberger, et al. Highly accurate protein structure prediction with AlphaFold. Nature 596(7873), 583–9 (2021).

[68] H.J.C. Berendsen, D. Van der Spoel, R. Van Drunen. GROMACS: A message-passing parallel molecular dynamics implementation. Comput. Phys. Commun. 91(1–3), 43–56 (1995).

[69] A.D. Mackerell, M. Feig, C.L. Brooks. Extending the treatment of backbone energetics in protein force fields: Limitations of gas‐phase quantum mechanics in reproducing protein conformational distributions in molecular dynamics simulations. J. Comput. Chem. 25(11), 1400–15 (2004).

[70] E. S. Istifli, P.A. Netz, A. Sihoglu Tepe, C. Sarikurkcu, B. Tepe. Understanding the molecular interaction of SARS-CoV-2 spike mutants with ACE2 (angiotensin converting enzyme 2). J Biomol Struct Dyn. 2022;40(23):12760-12771

[71] H. Woo, S.-J. Park, . K. Choi, et al. Developing a Fully Glycosylated Full-Length SARS-CoV-2 Spike Protein Model in a Viral Membrane. J. Phys. Chem. B 124, 33, 7128–7137 (2020).

[72] A.D. MacKerell, D. Bashford, M. Bellott, R.L. Dunbrack, J.D. Evanseck, M.J. Field, et al. All-Atom Empirical Potential for Molecular Modeling and Dynamics Studies of Proteins. J. Phys. Chem. B. 102(18), 3586–616 (1998).

[73] S. Piana, K. Lindorff-Larsen, D.E. Shaw. How Robust Are Protein Folding Simulations with Respect to Force Field Parameterization? Biophys J. 100(9), L47–9 (2011).

[74] K. Nakamoto. The Urey — Bradley Force Field: Its Significance and Application. In: Developments in Applied Spectroscopy. Boston, MA: Springer US; 1964. p. 158–68.

[75] L.H. Ngai, R.H. Mann. A transferable Urey-Bradley force field and the assignments of some mixed halomethanes.

[76] B. Hess, H. Bekker, H.J.C. Berendsen, J.G.E.M Fraaije. LINCS: A linear constraint solver for molecular simulations. J. Comput. Chem. 18(12), 1463–72 (1997).

[77]  G. Bussi, D. Donadio, M. Parrinello. Canonical sampling through velocity rescaling. J. Chem. Phys. 126(1), (2007).

[78]  M. Parrinello, A. Rahman. Polymorphic transitions in single crystals: A new molecular dynamics method. J. Appl. Phys. 52(12), 7182–90 (1982).

[79]  S. Nosé, M.L. Klein. Constant pressure molecular dynamics for molecular systems. Mol. Phys. 50(5), 1055–76 (1983).

[80]  T. Darden, D. York, L. Pedersen. Particle mesh Ewald: An $N \cdot \log(N)$ method for Ewald sums in large systems. J. Chem. Phys. 98(12), 10089–92 (1993).

[81]  G. Van Rossum, F.L. Drake. Python 3 Reference Manual. Scotts Valley, CA: CreateSpace; 2009.

[82]  W. Humphrey, A. Dalke, K. Schulten. VMD: Visual molecular dynamics. J. Mol. Graph. 14(1), 33–8 (1996).

[83]  M. Hebditch, M.A. Carballo-Amador, S. Charonis, R. Curtis, J. Warwicker. Protein–Sol: a web tool for predicting protein solubility from sequence. Bioinformatics 33, 3098–3100 (2017).

[84]  M. Hebditch, J. Warwicker. Web-based display of protein surface and pH-dependent properties for assessing the developability of biotherapeutics. Sci Rep 9, 1969 (2019)

[85]  L. Zhao, Z. Cao, Y. Bian, G. Hu, J. Wang, Y. Zhou. Molecular Dynamics Simulations of Human Antimicrobial Peptide LL-37 in Model POPC and POPG Lipid Bilayers. Int. J. Mol. Sci. 19(4), 1186 (2018).

[86]  N. Amdursky and M. M. Stevens. Circular Dichroism of Amino Acids: Following the Structural Formation of Phenylalanine. Chem. Phys. Chem. 16(13), 2768-2774 (2015)

[87]  N. Zaman, N. Parvaiz, F. Gul, et al. Dynamics of water-mediated interaction effects on the stability and transmission of Omicron. Sci. Rep. 13, 20894 (2023).

[88]  J. Singh, S. Vashishtha, S. A. Rahman, et al. Energetics of Spike Protein Opening of SARS-CoV-1 and SARS-CoV-2 and Its Variants of Concern: Implications in Host Receptor Scanning and Transmission. Biochemistry 61, 20, 2188–219 (2022).

[89]  N. Bhattacharjee and P. Biswas. Structural patterns in alpha helices and beta sheets in globular proteins. Protein and Peptide Letters. 16(8), 953-960, (2009).

[90]  N. Chitra, and C. L. Dias. Hydrophobic interactions and hydrogen bonds in β-sheet formation. The journal of chemical physics. 139(11), (2013).

[91]  F. Chiti. Relative importance of hydrophobicity, net charge, and secondary structure propensities in protein aggregation. Protein Misfolding, Aggregation, and Conformational

Diseases: Part A: Protein Aggregation and Conformational Diseases. Boston, MA: Springer US, 43-59 (2006).

[92] R. Funari, N. Bhalla, L. Gentile. Measuring the Radius of Gyration and Intrinsic Flexibility of Viral Proteins in Buffer Solution Using Small-Angle X-ray Scattering. ACS Meas. Sci. Au 2, 547−552 (2022).

[93] M. Gur, E. Taka, S.Z. Yilmaz, C. Kilinc, U. Aktas, M. Golcuk. Conformational transition of SARS-CoV-2 spike glycoprotein between its closed and open states. J. Chem. Phys. 153, 075101 (2020).

[94] R. Henderson, R.J. Edwards, K. Mansouri, K. Janowska, V. Stalls, S.M.C. Gobeil, M. Kopp, D. Li, R. Parks, A.L. Hsu, M.J. Borgnia, B.F. Haynes, P. Acharya. Controlling the SARS-CoV-2 spike glycoprotein conformation. Nat Struct Mol Biol 27, 925–933 (2020).

[95] M.H. Peters, O. Bastidas, D.S. Kokron, C.E. Henze. Static all-atom energetic mappings of the SARS-Cov-2 spike protein and dynamic stability analysis of "Up" versus "Down" protomer states. PLoS ONE 15(11), e0241168 (2020).

[96] J. Singh, S. Vashishtha, S. A. Rahman, et al. Energetics of Spike Protein Opening of SARS-CoV-1 and SARS-CoV-2 and Its Variants of Concern: Implications in Host Receptor Scanning and Transmission. Biochemistry 61, 20, 2188–219 (2022).

[97] N.J. Hardenbrook, P. Zhang. A structural view of the SARS-CoV-2 virus and its assembly. Current Opinion in Virology 52, 123–134 (2022).

[98] R. Yan, Y. Zhang, Y. Li, L. Xia, et al. Structural basis for the recognition of SARS-CoV-2 by full-length human ACE2. Science 367, 1444–1448 (2020).

[99] S. Lv, Y.-Q- Deng, Q. Ye, L. Cao, et al. Structural basis for neutralization of SARS-CoV-2 and SARS-CoV by a potent therapeutic antibody. Science 369, 1505–1509 (2020).

[100] Md.L. Hossen, P. Baral, T. Sharma, B. Gerstman, P. Chapagain. Significance of the RBD mutations in the SARS-CoV-2 omicron: from spike opening to antibody escape and cell attachment. Phys. Chem. Chem. Phys. 24, 9123-9129 (2022).

[101] Z. Zhao, J. Zhou, M. Tian, et al. Omicron SARS-CoV-2 mutations stabilize spike up-RBD conformation and lead to a non-RBM-binding monoclonal antibody escape. Nat. Commun. 13, 4958 (2022).

[102] L. Zeng, J. Li, M. Lv, Z. Li, L. Yao, J. Gao, Q. Wu, Z. Wang, X. Yang, G. Tang, G. Qu. G. Jiang. Environmental Stability and Transmissibility of Enveloped Viruses at Varied Animate and Inanimate Interfaces. Environ. Health 1, 1, 15–31 (2023).

[103] I. Samandoulgou, R. Hammami, R. Morales Rayas, I. Fliss, J. Jean. Stability of Secondary and Tertiary Structures of Virus-Like Particles Representing Noroviruses: Effects of pH, Ionic

Strength, and Temperature and Implications for Adhesion to Surfaces. Appl Environ Microbiol 81, (2015).

[104] D. Bao, C. Lu, T. Ma, G. Xu, Y. Mao, L. Xin, S. Niu, Z. Wu, X. Li, Q. Teng, Z. Li,Q. Liu. Hydrophobic Residues at the Intracellular Domain of the M2 Protein Play an Important Role in Budding and Membrane Integrity of Influenza Virus. J Virol 96:e00373-22 (2022).

[105] A.J. Prussin, D.O. Schwake, K. Lin, D.L. Gallagher, L. Buttling, L.C. Marr. Survival of the Enveloped Virus Phi6 in Droplets as a Function of Relative Humidity, Absolute Humidity, and Temperature. Appl Environ Microbiol 84, e00551-18 (2018).

[106] K. Lin, C.R. Schulte, L.C. Marr. Survival of MS2 and Φ6 viruses in droplets as a function of relative humidity, pH, and salt, protein, and surfactant concentrations. PLoS One. 15(12), e0243505 (2020)